\documentclass[runningheads, envcountsame, a4paper]{llncs}

\usepackage{enumitem}
\usepackage{bm}
\usepackage{xspace}
\usepackage{color}
\usepackage{mathtools}
\usepackage{subcaption}
\usepackage{booktabs}
\usepackage{multirow}
\usepackage{adjustbox}
\usepackage[ruled, vlined, linesnumbered]{algorithm2e}
\usepackage[font=small,labelfont=bf,tableposition=top]{caption}
\usepackage{tikz}
\usepackage{dsfont}
\usepackage{pifont}% http://ctan.org/pkg/pifont

\usepackage{amsthm}
\usepackage[bb=boondox]{mathalfa}
\usepackage{cite}
\usepackage[hyphens]{url}
\usepackage[misc]{ifsym}

\newcommand{\vpara}[1]{\vspace{0.08in}\noindent\textbf{#1}}
\newcommand{\hide}[1]{} %hide
\newcommand{\numberthis}{\addtocounter{equation}{1}\tag{\theequation}}
\newcommand\hmm[1]{\ifnum\ifhmode\spacefactor\else2000\fi>1000 \uppercase{#1}\else#1\fi}

\newcommand{\ie}{{\sl i.e.\xspace}}
\newcommand{\etc}{{\sl etc.}}

\newcommand{\trans}{^\top}

\newcommand{\mc}{\mathcal}
\newcommand{\norm}[1]{\left\lVert #1 \right\rVert}

\newcommand{\vc}{\mathbf}
\newcommand{\mat}{\mathbf}
\newcommand{\tnsr}{\mc}

\renewcommand{\phi}{\varphi}
\newcommand{\tst}[1]{\tnsr{X}^{(#1)}}
\newcommand{\tsid}{\tnsr{I}}
\newcommand{\ord}[1]{o({#1})}  % the order of the tensor #1
\newcommand{\matv}[2]{\mat{V}^{(#1)}_{#2}}  % matrix V for motif m index i
\newcommand{\vcmu}{\bm{\mu}}
\newcommand{\matvcs}[1]{\mat{V}^{*}_{#1}}  % consensus matrix V for the node type #1
\newcommand{\tmatv}{\widetilde{\mat{V}}}  % variable for auxiliary function
\newcommand{\subst}[1]{(#1)_{s,t}}  % variable for auxiliary function
\newcommand{\motifset}{\mc{M}}

\newcommand{\cmark}{\ding{51}}%
\newcommand{\xmark}{\ding{55}}%

\newcommand{\cmr}[1]{(\textcolor{red}{#1})}

\newcommand{\toaddback}{\hide} % To add back in camera-ready; change color to show

\newcommand{\fakeref}{}

\newcommand{\mochin}{\textsc{MoCHIN}\xspace}

\begin{document}
\title{User-Guided Clustering in Heterogeneous Information Networks via Motif-Based Comprehensive Transcription}

%\author{
%Yu Shi\footnotemark[1]\ \ \ \ 
%Xinwei He\footnotemark[1]\ \ \ \ 
%Naijing Zhang\footnotemark[1]\ \ \ \ 
%Carl Yang\ \ \ \ 
%Jiawei Han\\
%{\fontsize{10pt}{12pt}\selectfont{\text{University of Illinois at Urbana-Champaign, Urbana, IL USA}}}   \\
%{\fontsize{10pt}{12pt}\selectfont{\text{\{yushi2, xhe17, nzhang31, jiyang3, hanj\}@illinois.edu}}}\\
%}
\author{
Yu Shi$^{*}$ \and
Xinwei He$^{*}$ \and
Naijing Zhang$^{*}$ \and
Carl Yang \and
Jiawei Han 
}
\authorrunning{Y. Shi, et al.}
\titlerunning{User-Guided Clus. in HINs via Motif-Based Comprehensive Transcription}

\institute{University of Illinois at Urbana-Champaign, Urbana, IL USA\\
\email{\{yushi2, xhe17, nzhang31, jiyang3, hanj\}@illinois.edu}}

\tocauthor{Yu Shi, Xinwei He, Naijing Zhang, Carl Yang, and Jiawei Han}
\toctitle{User-Guided Clustering in Heterogeneous Information Networks via Motif-Based Comprehensive Transcription}

\maketitle

{
\renewcommand{\thefootnote}{\fnsymbol{footnote}}
\footnotetext[1]{These authors contributed equally to this work.}
}

%%%%%%%%%%%%%%%%%%%%%%%%%%%%%%%%%%%%%%%%%%%%%%%%%%%%%%
%% Abstract
%!TEX root = hin_motif_clus.tex

\begin{abstract}
Heterogeneous information networks (HINs) with rich semantics are ubiquitous in real-world applications. 
For a given HIN, many reasonable clustering results with distinct semantic meaning can simultaneously exist.
User-guided clustering is hence of great practical value for HINs where users provide labels to a small portion of nodes.
To cater to a broad spectrum of user guidance evidenced by different expected clustering results, carefully exploiting the signals residing in the data is potentially useful.
Meanwhile, as one type of complex networks, HINs often encapsulate higher-order interactions that reflect the interlocked nature among nodes and edges.
Network motifs, sometimes referred to as meta-graphs, have been used as tools to capture such higher-order interactions and reveal the many different semantics.
We therefore approach the problem of user-guided clustering in HINs with network motifs.
In this process, we identify the utility and importance of directly modeling higher-order interactions without collapsing them to pairwise interactions.
To achieve this, we comprehensively transcribe the higher-order interaction signals to a series of tensors via motifs and propose the \mochin model based on joint non-negative tensor factorization.
This approach applies to arbitrarily many, arbitrary forms of HIN motifs. %, which is often necessary for the application scenario in HINs due to their rich and diverse semantics encapsulated in the heterogeneity.
An inference algorithm with speed-up methods is also proposed to tackle the challenge that tensor size grows exponentially as the number of nodes in a motif increases.
We validate the effectiveness of the proposed method on two real-world datasets and three tasks, and \mochin outperforms all baselines in three evaluation tasks under three different metrics.
Additional experiments demonstrated the utility of motifs and the benefit of directly modeling higher-order information especially when user guidance is limited.
\footnote{The code and the data are available at \url{https://github.com/NoSegfault/MoCHIN}.}
\end{abstract}


\begin{figure}[t]
 \centering\includegraphics[width=.9\linewidth]{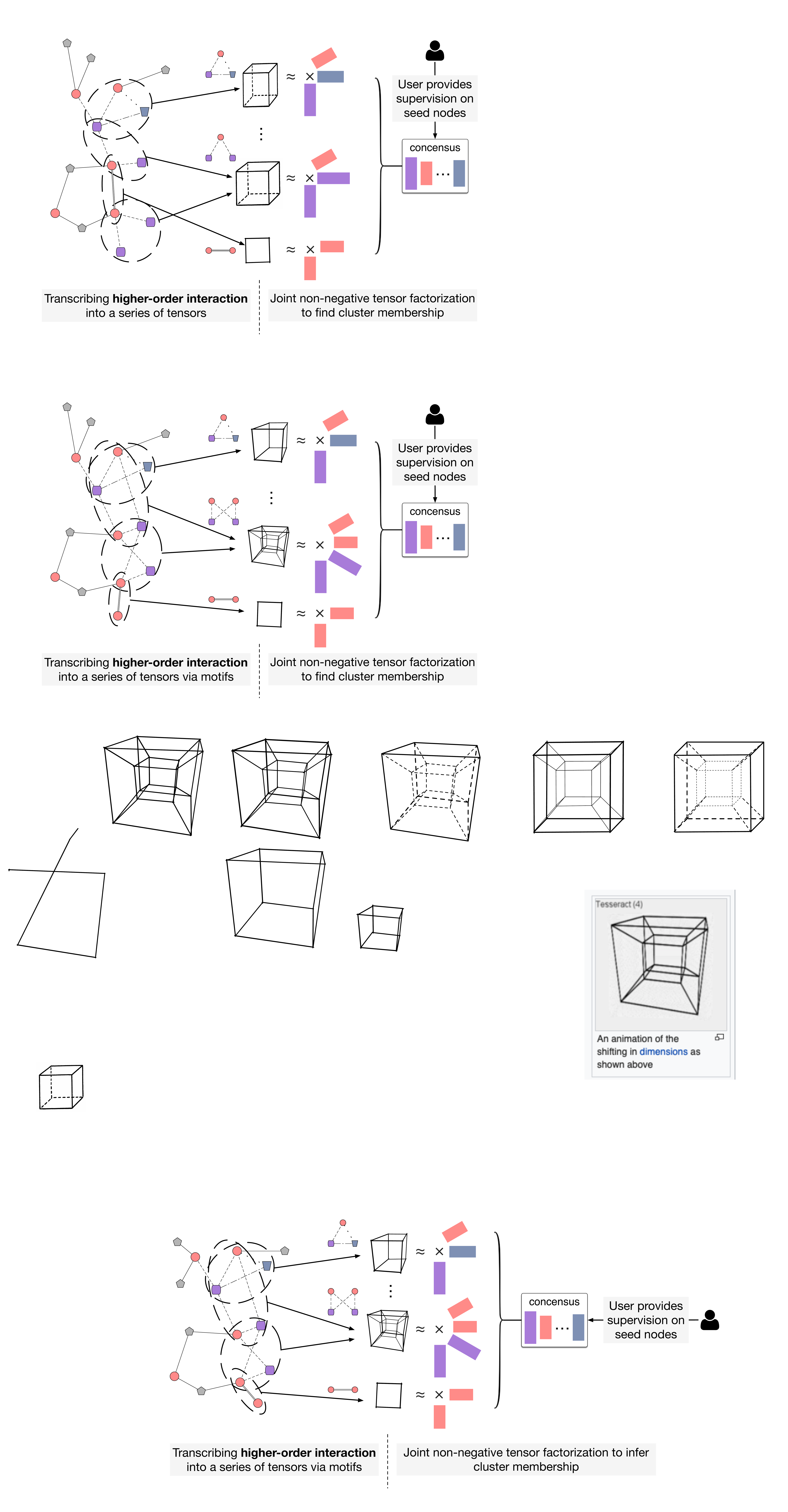}
 \caption[]{
 Overview of the proposed method \mochin that directly models all nodes in higher-order interactions where each type of nodes in the HIN corresponds to a color and a shape in the figure.
% To leverage signals from higher-order interactions without collapsing them into pairwise interactions, \mochin transcribes such information into a series of sparse tensors, which can provide a rich pool of fine-grained semantics and thereby fit a wider spectrum of guidance provided by different users.
 %The order of each tensor is identical to the number of nodes in the corresponding motif.
 %The tensors constructed in this way are sparse.
 %In the task of user-guided clustering in HINs, the tensors constructed as such can provide a rich pool of fine-grained semantics, which can thereby fit a wider spectrum of guidance provided by different users.
 }\label{fig::toy-motifs}
\end{figure}

\section{Introduction}\label{sec::introduction}
Heterogeneous information network (HIN) has been shown to be a powerful approach to model linked objects with informative type information~\cite{shi2017survey, sun2013mining, shi2017prep, shi2018aspem}.
%Many HIN-based methodologies have been proposed for applications such as classification, clustering, recommendation, and outlier detection~\cite{shi2017survey, sun2013mining}.
Meanwhile, the formation of complex networks is often partially attributed to the higher-order interactions among objects in real-world scenarios~\cite{benson2016higher, milo2002network, yaverouglu2014revealing}, where the ``players'' in the interactions are nodes in the network.
To reveal such higher-order interactions, researchers have since been using network motifs.
Leveraging motifs is shown to be useful in tasks such as clustering~\cite{yin2017local, benson2016higher}, ranking~\cite{zhao2018ranking} and representation learning~\cite{sankar2017motif}.
\footnote{
Higher-order interaction is sometimes used interchangeably with high-order interaction in the literature, and clustering using signals from higher-order interactions is referred to as higher-order clustering~\cite{yin2017local, benson2016higher}. \toaddback{~\cite{zhou2017local}}
Motifs in the context of HINs are sometimes called the meta-graphs, and we opt for motifs primarily because meta-graphs have been used under a different definition in the study of clustering~\cite{strehl2002cluster}.% as to be discussed in Section~\ref{sec::related-work}.
}

Clustering is a traditional and fundamental task in network mining~\cite{han2011data}. 
%\cmr{See note -- nicer ways to intro user-guided clustering with an intention to mention (1) the existence of not big data (no!) and (2) the need for indepth mining of semantics, especially given limited data.}
In the context of an HIN with rich semantics, reasonable clustering results with distinct semantic meaning can simultaneously exist. 
In this case, personalized clustering with user guidance can be of great practical value~\cite{sun2012integrating, shi2017survey, luo2014semi, jiang2017semi, gujral2018smacd}.
% the problem of user-guided clustering is particularly of interest, because HINs with different node and edge types can have multiple semantic facets and user guidance on the intention of clustering is often needed to generate more specific and meaningful results~\cite{sun2012integrating, shi2017survey, luo2014semi, jiang2017semi, gujral2018smacd}.
Carefully exploiting the fine-grained semantics in HINs via modeling higher-order interaction is a promising direction for such user-guided clustering since it could potentially generate a richer pool of subtle signals to better fit different users' guidance, especially when users cannot provide too much guidance and the supervision is hence weak.

%Exploiting higher-order interactions offers us the opportunity to better solve this important problem since it could potentially generate a richer pool of subtle signals to better fit different users' guidance.

However, it is non-trivial to develop a principled HIN clustering method that exploits signals revealed by motifs as comprehensively as possible.
This is because most network clustering algorithms are based on signals derived from the relatedness between each pair of nodes~\cite{han2011data}.
While a body of research has shown that it is beneficial for clustering methods to derive features for each node pair using motifs~\cite{huang2016meta, fang2016semantic, zhao2017meta, jiang2017semi, liu2018distance}, this approach essentially collapses a higher-order interaction into pairwise interactions, which is an irreversible process.
Such irreversible process is not always desirable as it could cause information loss.
For example, consider a motif instance involving three nodes -- A, B, and C.
After collapsing the higher-order interaction among A, B, and C into pairwise interactions, we are still able to sense the tie between A and C, but such a tie would no longer depend on B -- a potentially critical semantic facet of the relationship between A and C.
%The connection between this essential semantic facet and node B would be lost if only the collapsed pairwise interactions were modeled.
Such subtle information could be critical to distinguishing different user guidance. %, the information only accessible through the concrete node B can be critical in determining whether A and C should be clustered together.
We will further discuss this point by real-world example in Section~\ref{sec::motivation} and experiments in Section~\ref{sec::exp}.
Furthermore, although it is relatively easy to find semantically meaningful HIN motifs~\cite{huang2016meta, fang2016semantic}, motifs in HINs can have more complex topology compared to motifs in homogeneous networks do~\cite{benson2016higher, yin2017local}. 
In order to fully unleash the power of HIN motifs and exploit the signals extracted by them, we are motivated to propose a method that applies to arbitrary forms of HIN motifs.

To avoid such information loss with a method applicable to arbitrary forms of motifs, we propose to directly model the higher-order interactions by comprehensively transcribing them into a series of tensors.
As such, the complete information of higher-order interactions is preserved.
%no matter how complex the motifs are, where complex motifs are commonly seen in HINs.
Based on this intuition, we propose the \mochin model, short for \textbf{Mo}tif-based \textbf{C}lustering in \textbf{HIN}s, with an overview in Figure 1.
\mochin first transcribes information revealed by motifs into a series of tensors and then performs clustering by joint non-negative tensor decomposition with an additional mechanism to reflect user guidance.
%This approach does not rely on the pairwise clustering methods and can better retain the information captured by different motifs to suit the needs of user-guided clustering in the semantic-rich HINs.
%This approach does not rely on the pairwise clustering methods and can yield a rich pool of fine-grained signals and thereby fit a wider spectrum of guidance provided by different users.

%\cmr{Add a little more on hyper-edge in the traditional scenario, and may discuss further on comparison to pairwise conversion (against Benson's?).}
In this direction, an additional challenge arises from inducing tensor via corresponding motif -- the size of the tensor grows exponentially as the number of nodes involved in the motif increases.
Fortunately, motif instances are often sparse in real-world networks just as the number of edges is usually significantly smaller than the number of node pairs in a large real-world network.
This fact is to be corroborated in Section~3\fakeref{} of the supplementary file.
We hence develop an inference algorithm taking advantage of the sparsity of the tensors and the structure of the proposed \mochin model.
%Two real-world datasets and three tasks validate the effectiveness of the proposed model and the inference algorithm. 

Lastly, we summarize our contributions as follows: 
%\begin{enumerate}
%\item 
(i) we identify the utility of modeling higher-order interaction without collapsing it into pairwise interactions to avoid losing the rich and subtle information captured by motifs;
%\item
(ii) we propose the \mochin model that captures higher-order interaction via motif-based comprehensive transcription;
(iii) we develop an inference algorithm and speed-up methods for \mochin;
%\item
(iv) experiments on two real-world HINs and three tasks demonstrated the effectiveness of the proposed method as well as the utility of the tensor-based modeling approach in user-guided HIN clustering.
%\end{enumerate}

%!TEX root = hin_motif_clus.tex

\section{Related Work}\label{sec::related-work}
\vpara{Network motifs and motifs in HINs.}
%The formation of complex networks is often partially attributed to the higher-order interactions among objects in real-world scenarios~\cite{benson2016higher, milo2002network, yaverouglu2014revealing}.
%The modeling of higher-order interactions has been shown to be useful in many research areas such as neuroscience \cite{sporns2004motifs} and biological networks \cite{prvzulj2007biological}.%, and social networks \cite{ugander2013subgraph}. 
Network motifs, or graphlets, are usually used to identify higher-order interactions~\cite{yin2017local, benson2016higher, milo2002network, yaverouglu2014revealing}.
One popular research direction on network motifs has centered on efficiently counting motif instances such as triangles and more complex motifs \cite{ahmed2015efficient, stefani2017triest}.
Applications of motifs have also been found in tasks such as network partition and clustering~\cite{yin2017local, benson2016higher, li2017inhomogeneous, zhou2017local} as well as ranking~\cite{zhao2018ranking}.
\toaddback{Researchers have also studied enriching motif with additional attributes, such as temporal information, which has been shown to be instrumental in various network mining tasks~\cite{paranjape2017motifs, li2018temporal}.}

%\vpara{Motifs in heterogeneous information networks.}
In the context of HINs, network motifs are sometimes referred to as meta-graphs or meta-structures and have been studied recently \cite{sankar2017motif, huang2016meta, fang2016semantic, jiang2017semi, zhao2017meta, liu2018distance, liu2018interactive, yang2018meta}.
Many of these works study pairwise relationship such as relevance or similarity \cite{huang2016meta, fang2016semantic, zhao2017meta, liu2018distance, liu2018interactive}, and some other address the problem of representation learning \cite{sankar2017motif, yang2018meta} and graph classification \cite{yang2018node}.
Some of these prior works define meta-graphs or meta-structures to be directed acyclic graphs \cite{zhao2017meta, huang2016meta}, whereas we do not enforce this restriction on the definition of HIN motifs.
%We also remark that the term ``meta-graph'' is sometimes defined as a derived graph with indicator vectors being its vertices (nodes)~\cite{strehl2002cluster}, and clustering problem based on this definition of meta-graph has been studied for more than a decade~\cite{strehl2002cluster, mimaroglu2011combining, punera2007soft}.
%Therefore, we stick to the term ``motif'' to refer to the higher-order structural pattern of interest in this paper.

\vpara{Clustering in heterogeneous information networks.}
As a fundamental data mining problem, clustering has been studied for HINs~\cite{shi2017survey, sun2013mining, sun2009ranking, li2017semi, sun2012integrating, shi2014ranking}. 
One line of HIN clustering study leverages the synergetic effect of simultaneously tackling ranking and clustering~\cite{sun2009ranking, shi2014ranking}. \toaddback{chen2015clustering}
Clustering on specific types of HINs such as those with additional attributes has also been studied~\cite{li2017semi}. 
Wu et al.~\cite{wu2017tensor} resort to tensor for HIN clustering.
Their solution employs one tensor for one HIN and does not model different semantics implied by different structural patterns.

User guidance brings significantly more potentials to HIN clustering by providing a small portion of seeds~\cite{sun2012integrating, shi2017survey}, which enables users to inject intention of clustering.
To reveal the different semantics in an HIN, pioneering works exploit the meta-path, a special case of the motif, and reflect user-guidance by using the corresponding meta-paths~\cite{sun2012integrating, luo2014semi}.

%To exploit more complex structural patterns with greater distinctive power than that of meta-path, methods have been developed atop motifs or meta-graphs that satisfy certain constraints.
%there are no existing studies on motif-based HIN clustering applicable to arbitrarily many, arbitrary forms of HIN motifs.
To the best of our knowledge, a recent preprint~\cite{carranza2018higher} is the only paper that tackles HIN clustering and applies to arbitrary forms of HIN motifs, which is not specifically designed for the scenario with user guidance.
Given an HIN and a motif (\ie, typed-graphlet), this method filter the original adjacent matrix to derive the typed-graphlet adjacency matrix and then perform spectral clustering on the latter matrix.
While being able to filter out information irrelevant to the given motif, this method essentially exploits the edge-level pairwise information in the adjacent matrix rather than directly modeling each occurrence of higher-order interaction. 
Other related works include a meta-graph--guided random walk algorithm~\cite{jiang2017semi}, which is shown to outperform using only meta-paths. %, which requires the used motifs to have undirected edges. % to ensure symmetry in the transition matrix.
Note that this method cannot distinguish motif AP4TPA from meta-path APTPA, which are to be introduced in Section~\ref{sec::motivation}. %, due to the design on how a random walk is sampled under a motif.
Sankar et al.~\cite{sankar2017motif} propose a convolutional neural network method based on motifs which can potentially be used for user-guided HIN clustering. 
This approach restricts the motifs of interest to those with a target node, a context node, and auxiliary nodes.
Gujral et al.~\cite{gujral2018smacd} propose a method based on tensor constructed from stacking a set of adjacency matrices, which can successfully reflect user guidance and different semantic aspects.
This method essentially leverages features derived for node pairs. %, instead of directly modeling higher-order interactions among multiple nodes.

\toaddback{Add Pan Li's work in camera-ready.}

We additionally review the related work on matrix and tensor factorization for clustering  in the supplementary file for this paper.
These studies are relevant but cannot be directly applied to the scenario of higher-order HIN clustering.

%Due to space limitations, we do not further discuss related work on tensor factorization in this section.
%We will provide necessary background and reference when describing the proposed model.

%!TEX root = hin_motif_clus.tex

\section{Preliminaries}\label{sec::prob-def}
In this section, we define related concepts and notations. 
%We first introduce the definition of heterogeneous information networks~\cite{sun2013mining}.

\begin{definition}[Heterogeneous information network and schema ~\cite{sun2013mining}]
An {information network} is a directed graph $G = (\mc{V}, \mc{E})$ with a node type mapping $\phi: \mc{V} \rightarrow \mc{T}$ and an edge type mapping $\psi: \mc{E} \rightarrow \mc{R}$. 
When $|\mc{T}| > 1$ or $|\mc{R}| > 1$, the network is referred to as a \textbf{heterogeneous information network} (\textbf{HIN}).
The \textbf{schema} of an HIN is an abstraction of the meta-information of the node types and edge types of the given HIN.
\end{definition}

\begin{figure}[t]
  \centering
  \begin{subfigure}[t]{0.24\linewidth}
    \centering\includegraphics[width=\linewidth]{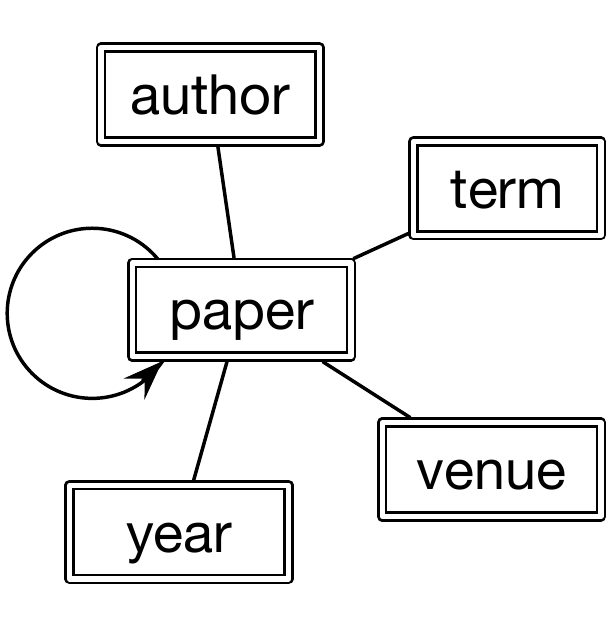}
    \caption{The schema~\cite{shi2018easing}.}\label{fig::dblp-schema}
  \end{subfigure}
  \qquad  
  \begin{subfigure}[t]{0.31\linewidth}
    \centering\includegraphics[width=\linewidth]{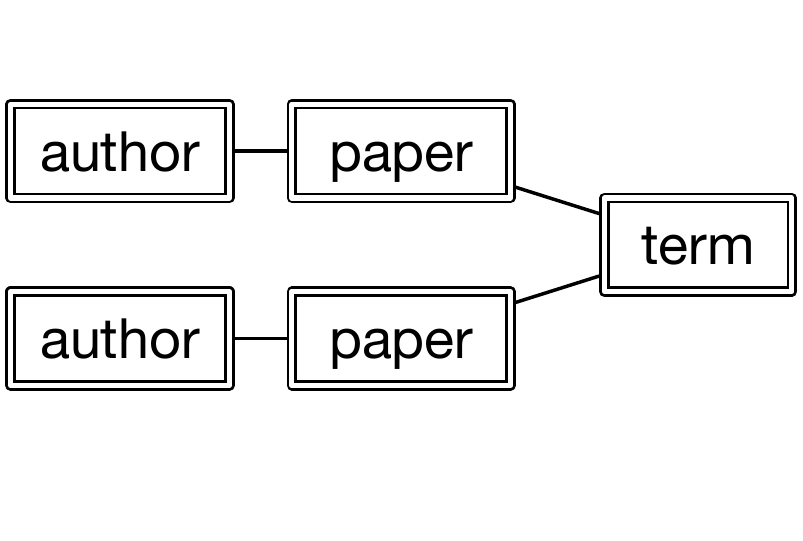}
    \caption{The APTPA motif, which is also a meta-path.}\label{fig::aptpa}
  \end{subfigure}
  \qquad
  \begin{subfigure}[t]{0.31\linewidth}
    \centering\includegraphics[width=\linewidth]{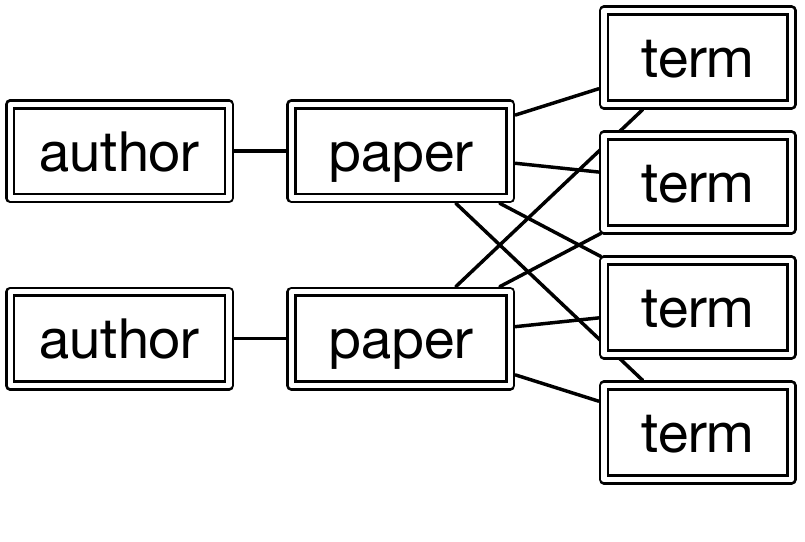}
    \caption{The AP4TPA motif.}\label{fig::ap4tpa}
  \end{subfigure}
  \caption{Examples of schema and motif in the DBLP network.}\label{fig::instance}
\end{figure}

As an example, Figure~\ref{fig::dblp-schema} illustrates the schema of the DBLP network to be used in Section~\ref{sec::exp}. We denote all nodes with the same type $t \in \mc{T}$ by $\mc{V}_t$. 

\begin{definition}[HIN motif and HIN motif instance]
In an HIN $G = (\mc{V}, \mc{E})$, an \textbf{HIN motif} is a structural pattern defined by a graph on the type level with its node being a node type of the original HIN and an edge being an edge type of the given HIN.
Additional constraints can be optionally added such as two nodes in the motif cannot be simultaneously matched to the same node instance in the given HIN.
Further given an HIN motif, an \textbf{HIN motif instance} under this motif is a subnetwork of the HIN that matches this pattern.
\end{definition}

%As discussed in Section~\ref{sec::related-work}, the motifs in HINs are sometimes referred to as the meta-graphs, and we choose motifs over meta-graphs in this paper primarily because meta-graph has been used under a different definition in the study of network clustering~\cite{strehl2002cluster, mimaroglu2011combining, punera2007soft}. 
Figure~\ref{fig::ap4tpa} gives an example of a motif in the DBLP network with four distinct terms referred to as $AP4TPA$.
If a motif is a path graph, it is also called a meta-path~\cite{sun2012integrating}.
The motif, $APTPA$, in Figure~\ref{fig::aptpa} is one such example.

\begin{definition}[Tensor, $k$-mode product, mode-$k$ matricization~\cite{papalexakis2017tensors}]
A \textbf{tensor} is a multidimensional array. 
For an $N$-th--order tensor $\tnsr{X} \in \mathbb{R}^{d_1 \times \ldots \times d_N}$, we denote its $(j_1, \ldots, j_N)$ entry by $\tnsr{X}_{j_1, \ldots, j_N}$.
The \textbf{$k$-mode product} of $\tnsr{X}$ and a matrix $ \mat{A} \in \mathbb{R}^{d_k \times d}$ is denoted by $\tnsr{Y} = \tnsr{X} \times_{k} \mat{A}$, 
where $\tnsr{Y} \in \mathbb{R}^{d_1 \times \ldots \times d_{k-1} \times d \times d_{k+1} \times \ldots \times d_N}$,
and $\tnsr{Y}_{\ldots, j_{k-1}, j, j_{k+1}, \ldots} = \sum_{s = 1}^{d_k} \tnsr{X}_{\ldots, j_{k-1}, s, j_{k+1}, \ldots}  \mat{A}_{s, j}$.
We denote matrix $\tnsr{X}_{(k)} \in \mathbb{R}^{(d_1 \cdot \ldots \cdot d_{k-1} \cdot d_{k+1} \cdot \ldots \cdot d_N) \times d_k}$ the \textbf{mode-$k$ matricization}, \ie, mode-$k$ unfolding, of the tensor $\tnsr{X}$, where the $i$-th column of $\tnsr{X}_{(k)}$ is obtained by vectorizing the $(n-1)$-th order tensor $\tnsr{X}_{\ldots, :, j, :, \ldots}$ with $j$ on the $k$-th index.
\end{definition}

For simplicity, we denote $\tnsr{X} \times_{i=1}^N \mat{A}_i \coloneqq \tnsr{X} \times_{1} \mat{A}_1 \times_{2} \ldots \times_{N} \mat{A}_N$. 
Additionally, we define $[\otimes_{i=1}^{N \backslash k} \mat{A}_i ] \coloneqq \mat{A}_1 \otimes \ldots \otimes \mat{A}_{k-1} \otimes \mat{A}_{k+1} \otimes \ldots \otimes \mat{A}_N$, where $\otimes$ is the Kronecker product~\cite{papalexakis2017tensors}.

Lastly, we introduce a useful lemma that converts the norm of the difference between two tensors to that between two matrices.
\begin{lemma}[\cite{de2000multilinear} ]\label{lem::unfold}
For all $k \in \{1, 2, \ldots, N\}$,
\begin{small}
\begin{align*}
& \norm{ \tnsr{X} - \tnsr{Y} \times_{i=1}^N \mat{A}_i }_F  = \norm{ \tnsr{X}_{(k)} - \mat{A}_k \tnsr{Y}_{(k)} [\otimes_{i=1}^{N \backslash k} \mat{A}_i ]\trans  }_F,
\end{align*}
\end{small}
where $\norm{\cdot}_F$ is the Frobenius norm.
\end{lemma}

%!TEX root = hin_motif_clus.tex

\begin{figure}[t]
 \centering\includegraphics[width=0.7\linewidth]{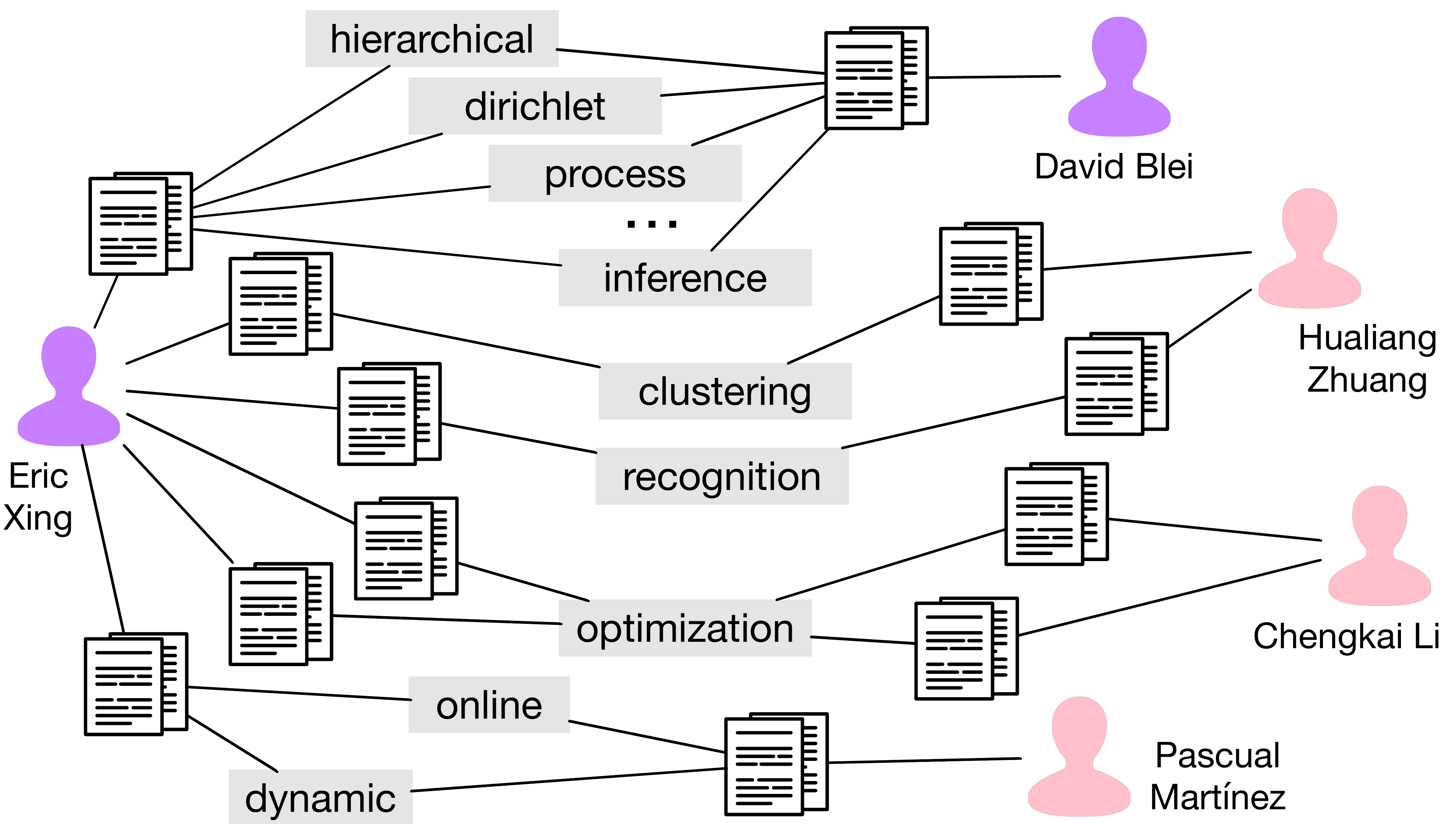}
 \caption[]{
A subnetwork of DBLP. % containing \textit{Eric Xing}, \textit{David Blei}, \textit{Hualiang Zhuang}, \textit{Chengkai Li}, and \textit{Pascual Martinez}. 
According to the ground truth data, \textit{Eric Xing} and \textit{David Blei} were graduated from the same research group. 
%While many path instances can be observed between \textit{Eric Xing} and authors from other groups, $AP4TPA$ motif instances can only be found between \textit{Eric Xing} and \textit{David Blei}. 
%Moreover, the concrete terms and papers involved in these motif instances are also informative if users wish to cluster authors from the same research group together.
 }\label{fig::subnetwork}
\end{figure}

\section{Higher-Order Interaction in Real-World Dataset.}\label{sec::motivation}
In this section, we use a real-world example to motivate the design of our method that aims to comprehensively model higher-order interactions revealed by motifs.

DBLP is a bibliographical network in the computer science domain~\cite{tang2008arnetminer} that contains nodes with type author, paper, term, \etc{} 
In Figure~\ref{fig::subnetwork}, we plot out a subnetwork involving five authors: \textit{Eric Xing}, \textit{David Blei}, \textit{Hualiang Zhuang}, \textit{Chengkai Li}, and \textit{Pascual Martinez}.
According to the ground truth labels, \textit{Xing} and \textit{Blei} graduated from the same research group, while the other three authors graduated from other groups.
Under meta-path $APTPA$, one would be able to find many path instances from \textit{Xing} to authors from different groups.
However, if we use motif $AP4TPA$, motif instances can only be found over \textit{Xing} and \textit{Blei}, but not between \textit{Xing} and authors from other groups.
This implies motifs can provide more subtle information than meta-paths do, and if a user wishes to cluster authors by research groups, motif $AP4TPA$ can be informative.

More importantly, if we look into the $AP4TPA$ motif instances that are matched to \textit{Xing} and \textit{Blei}, the involved terms such as \textit{dirichlet} are very specific to their group's research interest.
In other words, \textit{dirichlet} represents an important semantic facet of the relationship between \textit{Xing} and \textit{Blei}.
Modeling the higher-order interaction among \textit{dirichlet} and other nodes can therefore kick in more information. % even.% if users ultimately only wish to obtain clustering results on authors.
If one only used motifs to generate pairwise or edge-level signals, such information would be lost. 
In Section~\ref{sec::exp}, we will further quantitatively validate the utility of comprehensively modeling higher-order interactions.

%!TEX root = hin_motif_clus.tex

\section{The \mochin Model}\label{sec::modeling}
In this section, we describe the proposed model with an emphasis on its intention to comprehensively model higher-order interaction while availing user guidance.

%\subsection{Revisit on Clustering by Non-Negative Matrix Factorization}
\vpara{Revisit on clustering by non-negative matrix factorization.}
Non-negative matrix factorization (NMF) has been a popular method clustering~\cite{liu2013multi, lee2001algorithms}.
Usually with additional constraints or regularization, the basic NMF-based algorithm solves the following optimization problem for given adjacency matrix $M$ 
\vspace{-6pt}
\begin{small}
\begin{equation}\label{eq::nmf}
\min_{ \mat{V}_1, \mat{V}_2 \geq 0 } \norm{\mat{M} - \mat{V}_{1}\trans \mat{V}_{2}}_F^2,
\end{equation}\vspace{-6pt}
\end{small}\\
where $\norm{\cdot}_F$ is the Frobenius norm, $\mat{A} \geq 0$ denotes matrix $\mat{A}$ is non-negative, and $\mat{V}_1$, $\mat{V}_2 $ are two $C \times |\mc{V}|$ matrices with $C$ being the number of clusters. 
In this model, the $j$-th column of $\mat{V}_1$ or that of $\mat{V}_2$ gives the inferred cluster membership of the $j$-th node in the network.
%The intuition of the model stems from using the inner product of the cluster membership of the $j_1$-th node and that of the $j_2$-th node to reconstruct the existence of the edge represented by non-zero entry $(j_1, j_2)$ in the adjacent matrix.

%\subsection{Single-Motif--Based Clustering in HINs}
\vpara{Single-motif--based clustering in HINs.}
Recall that an edge essentially characterizes the pairwise interaction between two nodes.
To model higher-order interaction without collapsing it into pairwise interactions, a natural solution to clustering is using the inferred cluster membership of all involved nodes to reconstruct the existence of each motif instance.
This solution can be formulated by non-negative tensor factorization (NTF), and studies on NTF per se and clustering via factorizing a single tensor can be found in the literature~\cite{papalexakis2017tensors}. % may add back or not ~\cite{benson2015tensor, cao2015robust}

Specifically, given a single motif $m$ with $N$ nodes having node type $t_1$, \ldots, $t_N$ of the HIN, we transcribe the higher-order interaction revealed by this motif to a $N$-th--order tensor $\tnsr{X}$ with dimension $|\mc{V}_{t_1}| \times \ldots \times |\mc{V}_{t_N}|$.
We set the $(j_1, \ldots, j_N)$ entry of $\tnsr{X}$ to $1$ if a motif instance exists over the following $n$ nodes: $j_1$-th of $\mc{V}_{t_1}$,  \ldots, $j_N$-th of $\mc{V}_{t_N}$; and set it to $0$ otherwise.
By extending Eq.~\eqref{eq::nmf}, whose objective is equivalent to $\norm{\mat{M} - \mat{V}_{1}\trans \mat{I} \mat{V}_{2}}_F^2$ with $\mat{I}$ being the identity matrix, we can approach the clustering problem by solving
\vspace{-6pt}
\begin{small}
\begin{equation}
\min_{ \mat{V}_1, \mat{V}_2 \geq 0 } \norm{\tnsr{X} - \tsid \times_{i=1}^{{N}} \mat{V}_{i}}_F^2 + \lambda \sum_{i = 1}^{N} \norm{\mat{V}_{i}}_1, 
\end{equation}\vspace{-6pt}
\end{small}\\
where $\tsid$ is the $N$-th order identity tensor with dimension $C \times \ldots \times C$, $\norm{\cdot}_1$ is the entry-wise $l$-1 norm introduced as regularization to avoid trivial solution, and $\lambda$ is the regularization coefficient.
We also note that this formulation is essentially the CP decomposition~\cite{papalexakis2017tensors} with additional l-$1$ regularization and non-negative constraints.
We write this formula in a way different from its most common form for notation convenience in the inference section (Section~\ref{sec::inference}) considering the presence of regularization and constraints.

%\begin{table}[t!]
%\centering
%%\resizebox{0.48\textwidth}{!}{
%\begin{tabular}{| c | c |}
%\hline
%\textbf{Symbol}  & \textbf{Definition} \\ 
%\hline \hline
%$\mc{V}$, $\mc{E}$ & The set of nodes and the set of edges \\
%$\mc{T}$, $\mc{R}$, $\mc{M}$ & The set of node types, edge types, and candidate motifs \\
%$\phi$, $\psi$ & The node type mapping and the edge type mapping \\
%$\mc{V}_t$ & The set of all nodes with type $t$ \\
%$o(m)$ & The number of nodes in motif $m \in \mc{M}$ \\
%\hline
%$\tst{m}$ & The tensor constructed from motif $m$ \\
%$\mat{M}^{(t)}$ & The seed mask matrix for node type $t$ \\
%$\matv{m}{i}$ & The cluster membership matrix   for the $i$-th node in motif $m$ \\
%$\matvcs{t}$ & The consensus matrix for node type $t$\\
%$\vcmu$ & The vector $(\mu_1, \ldots, \mu_{|\mc{M}|})$ of motif weights \\
%\hline
%$C$ & The number of clusters \\
%$\lambda$, $\theta$, $\rho$ & The hyperparameters \\
%\hline
%$\times_k$ & The mode-k product of a tensor and a matrix\\
%$\otimes$ & The Kronecker product of two matrices\\
%\hline
%\end{tabular}
%%}
%\caption{Summary of symbols}\label{tab::symbol}
%%\vspace{-12pt}
%\end{table}

\begin{table}[t!]
\centering
\resizebox{\textwidth}{!}{
\begin{tabular}{| c | c || c | c |}
\hline
\textbf{Symbol}  & \textbf{Definition} & \textbf{Symbol}  & \textbf{Definition} \\ 
\hline \hline
$\mc{V}$, $\mc{E}$ & The set of nodes and the set of edges & $\tst{m}$ & The tensor constructed from motif $m$\\
$\phi$, $\psi$ & The node and the edge type mapping & $\mat{M}^{(t)}$ & The seed mask matrix for node type $t$ \\
\multirow{2}{*}{$\mc{T}$, $\mc{R}$, $\mc{M}$} & The set of node types, edge types,  & \multirow{2}{*}{$\matv{m}{i}$} & The cluster membership matrix \\
& and candidate motifs & &  for the $i$-th node in motif $m$\\
$\mc{V}_t$ & The set of all nodes with type $t$ & $\matvcs{t}$ & The consensus matrix for node type $t$ \\
$o(m)$ & The number of nodes in motif $m \in \mc{M}$ &  $\vcmu$ & The vector $(\mu_1, \ldots, \mu_{|\mc{M}|})$ of motif weights\\
\hline
$C$ & The number of clusters & $\times_k$ & The mode-k product of a tensor and a matrix \\
$\lambda$, $\theta$, $\rho$ & The hyperparameters & $\otimes$ & The Kronecker product of two matrices \\
\hline
\end{tabular}
}
\caption{Summary of symbols}\label{tab::symbol}
%\vspace{-12pt}
\end{table}

%\subsection{Proposed Model for Motif-Based Clustering in HINs}
\vpara{Proposed model for motif-based clustering in HINs.}
Real-world HINs often contain rich and diverse semantic facets due to its heterogeneity~\cite{sun2012integrating, shi2018easing, sun2013mining}.
To reflect the different semantic facets of an HIN, a set $\mc{M}$ of more than one candidate motifs are usually necessary for the task of user-guided clustering.
With additional clustering seeds provided by users, the \mochin model selects the motifs that are both meaningful and pertinent to the seeds.

To this end, we assign motif-specific weights $\vcmu = (\mu_1, \ldots, \mu_{|\mc{M}|})$, such that $\sum_{m \in \mc{M}} \mu_m = 1$ and $\mu_m \geq 0$ for all $m \in \mc{M}$.
Denote $\tst{m}$ the tensor for motif $m$, $\matv{m}{i}$ the cluster membership matrix for the $i$-th node in motif $m$, $o(m)$ the number of nodes in motif $m$, and $\phi(m, i)$ the node type of the $i$-th node in motif $m$.
For each node type $t \in \mc{T}$, we put together cluster membership matrices concerning this type and motif weights to construct the consensus matrix 
$
\matvcs{t} \coloneqq \sum_{ \phi(m,i) = t} \frac{\mu_m \matv{m}{i} }{ \sum_{i'=1}^{\ord{m}} {\mathds{1}}_{[\phi(m, i') = \phi(m,i) ]}},
$
where $\mathds{1}_{[P]}$ equals to $1$ if $P$ is true and 0 otherwise.
With this notation, $\sum_{i'=1}^{\ord{m}} {\mathds{1}}_{[\phi(m, i') = \phi(m,i) ]}$ is simply the number of nodes in motif $m$ that are of type $t$.

Furthermore, we intend to let 
(i) each cluster membership $\matv{m}{i}$ be close to its corresponding consensus matrix $\matvcs{\phi(m, i)}$ and 
(ii) the consensus matrices not assign seed nodes to the wrong cluster.
We hence propose the following objective with the third and the fourth term modeling the aforementioned two intentions
\vspace{-6pt}
\begin{small}
\begin{align*}
\mc{O} & = \sum_{m \in \motifset}{\norm{\tst{m} - \tsid^{(m)}\times_{i=1}^{\ord{m}} \matv{m}{i}}_F^2}  + \lambda \sum_{m \in \motifset}\sum_{i = 1}^{\ord{m}} \norm{\matv{m}{i}}_1\\
&  + \theta \sum_{m \in \motifset}\sum_{i = 1}^{\ord{m}}{\norm{\matv{m}{i} - \matvcs{\phi(m, i)} }_{F}^{2}} + \rho \sum_{t \in \mathcal{T}}{\norm{\mat{M}^{(t)} \circ \matvcs{t} }_{F}^{2}}, \numberthis\label{eq::obj}\vspace{-6pt}
\end{align*}
\end{small}\\
where $\circ$ is the Hadamard product and $\mat{M}^{(t)}$ is the seed mask matrix for node type $t$. 
Its $(i, c)$ entry $\mat{M}^{(t)}_{i, c} = 1$ if the $i$-th node of type $t$ is a seed node and it should not be assigned to cluster $c$, and $\mat{M}^{(t)}_{i, c} = 0$ otherwise.

Finally, solving the problem of HIN clustering by modeling higher-order interaction and automatically selecting motifs is converted to solving the following optimization problem with $\Delta$ being the standard simplex
\vspace{-6pt}
\begin{small}
\begin{equation}\label{eq::opt}
\min_{{ \{\matv{m}{i} \geq 0 \}, \vcmu \in \Delta}} \mc{O}.
\end{equation}\vspace{-6pt}
\end{small}\\
%To the best of our knowledge, there is no method similar to ours that simultaneously model multiple motifs in an HIN without decomposing higher-order interactions into pairwise interactions.

%!TEX root = hin_motif_clus.tex

\section{The Inference Algorithm}\label{sec::inference}
In this section, we first describe the algorithm for solving the optimization problem as in Eq.~\eqref{eq::opt}. 
Then, a series of speed-up tricks are introduced to circumvent the curse of dimensionality, so that the complexity is governed no longer by the dimension of the tensors but by the number of motif instances in the network. %where direct computation on the tensors would be problematic since a motif involving many nodes would induce a tensor with a formidably large number of entries.

%\subsection{Update $\matv{l}{k}$ and $\mathbf{\mu}$}
\vpara{Update $\matv{l}{k}$ and $\mathbf{\mu}$.}
Each clustering membership matrix $\matv{l}{k}$ with non-negative constraints is involved in all terms of the objective function (Eq.~\eqref{eq::obj}), where $l \in \mc{M}$ and $k \in \{1, \ldots, o(l)\}$.
We hence develop multiplicative update rules for $\matv{l}{k}$ that guarantees monotonic decrease at each step, accompanied by projected gradient descent (PGD) to find global optimal of $\vcmu = [\mu_1, \ldots, \mu_{|\mc{M}|}]\trans$. % by exploiting its convexity.
Overall, we solve the optimization problem by alternating between $\{\matv{l}{k}\}$ and $\vcmu$.

To update $\matv{l}{k}$ when $\{\matv{m}{i}\}_{(m, i) \neq (l, k)}$ and $\vcmu$ are fixed under non-negative constraints, we derive the following theorem.
For notation convenience, we further denote $\matvcs{t} = \sum_{ \phi(m,i) = t} \eta^{m}_i \matv{m}{i}$, where  $\eta^{m}_i \coloneqq \frac{\mu_{m}  }{ \sum_{i'=1}^{\ord{m}} {\mathds{1}}_{[\phi(m, i') = \phi(m,i) ]}  } $. 

\begin{theorem}\label{thm::update-v}
The following update rule for $\matv{l}{k}$ monotonically decreases the objective function.
\begin{small}
%\begin{align*}
%& \matv{l}{k}  \leftarrow 
% \matv{l}{k} \circ \Bigg[ 
% \frac{\tst{l}_{(k)}  [\otimes_{i=1}^{\ord{l} \backslash k} \matv{l}{i} ] \tsid^{(l)\top}_{(k)}}
% {\matv{l}{k} \tsid^{(l)}_{(k)} [\otimes_{i=1}^{\ord{l} \backslash k} \matv{l}{i} ]\trans [\otimes_{i=1}^{\ord{l} \backslash k} \matv{l}{i} ] \tsid^{(l)\top}_{(k)} } \\
%& \frac{ + \theta \eta^{l}_{k} \sum_{\phi(m,i) = \phi(l,k) }^{ (m,i) \neq (l,k)} [\matv{m}{i} - \matvcs{\phi(l, k)} + \eta^{l}_{k} \matv{l}{k}]^{+} }
%{ + \theta \eta^{l}_{k} \sum_{\phi(m,i) = \phi(l,k) }^{ (m,i) \neq (l,k)} ([\matv{m}{i} - \matvcs{\phi(l, k)} + \eta^{l}_{k} \matv{l}{k}]^{-} + \eta^{l}_{k} \matv{l}{k} ) } \\
%& \frac{  + \theta (1 - \eta^{l}_{k}) (\matvcs{\phi(l, k)} - \eta^{l}_{k} \matv{l}{k})}{+ \theta (1 - \eta^{l}_{k})^2 \matv{l}{k} + \rho  \eta^{l}_{k} {\mat{M}^{\phi(l,k)} \circ \matvcs{\phi(l,k)} + \lambda}} 
%\Bigg]^{\frac{1}{2}}, \numberthis\label{eq::update-v}
%\end{align*}
\begin{align*}
& \matv{l}{k}  \leftarrow 
 \matv{l}{k} \circ \Bigg[ 
 \frac{\tst{l}_{(k)}  [\otimes_{i=1}^{\ord{l} \backslash k} \matv{l}{i} ] \tsid^{(l)\top}_{(k)}         + \theta (1 - \eta^{l}_{k}) (\matvcs{\phi(l, k)} - \eta^{l}_{k} \matv{l}{k})}
 {\matv{l}{k} \tsid^{(l)}_{(k)} [\otimes_{i=1}^{\ord{l} \backslash k} \matv{l}{i} ]\trans [\otimes_{i=1}^{\ord{l} \backslash k} \matv{l}{i} ] \tsid^{(l)\top}_{(k)}     + \rho  \eta^{l}_{k} {\mat{M}^{\phi(l,k)} \circ \matvcs{\phi(l,k)}}} \\
& \frac{ + \theta \eta^{l}_{k} \sum_{\phi(m,i) = \phi(l,k) }^{ (m,i) \neq (l,k)} [\matv{m}{i} - \matvcs{\phi(l, k)} + \eta^{l}_{k} \matv{l}{k}]^{+} }
{ + \theta \eta^{l}_{k} \sum_{\phi(m,i) = \phi(l,k) }^{ (m,i) \neq (l,k)} ([\matv{m}{i} - \matvcs{\phi(l, k)} + \eta^{l}_{k} \matv{l}{k}]^{-} + \eta^{l}_{k} \matv{l}{k} )   + \theta (1 - \eta^{l}_{k})^2 \matv{l}{k}    + \lambda} 
\Bigg]^{\frac{1}{2}}, \numberthis\label{eq::update-v}
\end{align*}
\end{small}
where for any matrix $\mat{A}$, $[\mat{A}]^{+} \coloneqq \frac{|\mat{A}| + \mat{A}}{2}$, $[\mat{A}]^{-} \coloneqq \frac{|\mat{A}| - \mat{A}}{2}$.
\end{theorem}
We defer the proof of this theorem to Section~1\fakeref{} of the supplementary file.
For fixed $\{\matv{m}{i}\}$, the objective function Eq.~\eqref{eq::obj} is convex with respect to $\vcmu$. 
We therefore use PGD to update $\vcmu$, where the gradient can be analytically derived with straightforward calculation.%, which we omit due to space limitations.

\begin{algorithm}[t!]
\DontPrintSemicolon
\SetKwInOut{Input}{Input}
\SetKwInOut{Output}{Output}
\Input{\{$\tst{m}\}$, supervision $\mat{M}^{(t)}$, the number of clusters $C$, hyperparameters $\theta$, $\rho$, and $\lambda$}
\Output{the cluster membership matrices $\{\matvcs{t}\}$}
\Begin{
  \While{not converged}{
      \For{$m \in \mc{M}$}{
          \While{not converged}{
              \For{$i \in \{1, \ldots, o(m)\}$}{
                Find local optimum of $\matv{m}{i}$ by Eq.~\eqref{eq::update-v}.
              }
          }
      }
      Find global optimum of $\vcmu$ by PGD.
    }
}
\caption{The \mochin inference algorithm}\label{alg::inference}
\end{algorithm}

%\subsection{Computational Speed-Up}\label{sec::speed-up}
\vpara{Computational Speed-Up.}
%In this section, we describe a series of speed-up tricks, with which the complexity would be governed no longer by the dimension of the tensors but by the number of motif instances in the network.
Unlike the scenario where researchers solve the NTF problem with tensors of order independent of the applied dataset, our problem is specifically challenging because the tensor size grows exponentially with the tensor order.
For instance, the AP4TPA motif discussed in Section~\ref{sec::motivation} is one real-world example involving $8$ nodes, which leads to an $8$-th order tensor.
%Consequently, the fact that the tensor size grows exponentially with the order of the tensor poses a special challenge to conducting motif-based clustering via tensor factorization. 

In the proposed inference algorithm, the direct computation of three terms entails complexity subject to the size of the tensor: (i) the first term in the numerator of Eq.~\eqref{eq::update-v}, $\tst{l}_{(k)}  [\otimes_{i=1}^{\ord{l} \backslash k} \matv{l}{i} ] \tsid^{(l)\top}_{(k)}$, (ii) the first term in the denominator of Eq.~\eqref{eq::update-v}, $\matv{l}{k} \tsid^{(l)}_{(k)} [\otimes_{i=1}^{\ord{l} \backslash k} \matv{l}{i} ]\trans $ $ [\otimes_{i=1}^{\ord{l} \backslash k} \matv{l}{i} ] \tsid^{(l)\top}_{(k)}$, and (iii) the first term of the objective function Eq.~\eqref{eq::obj}, $\norm{\tst{m} - \tsid^{(m)}\times_{i=1}^{\ord{m}} \matv{m}{i}}_{F}^2$.
Fortunately, all these terms can be significantly simplified by exploiting the composition of dense matrix $[\otimes_{i=1}^{\ord{l} \backslash k} \matv{l}{i} ] \tsid^{(l)\top}_{(k)}$ and the sparsity of tensor $\tst{l}$ ($\tst{m}$).

Consider the example that motif $l \in \mc{M}$ involves $5$ nodes, each node type has $10,000$ node instances, and the nodes are to be clustered into $10$ clusters.
Then the induced dense matrix $[\otimes_{i=1}^{\ord{l} \backslash k} \matv{l}{i} ] \tsid^{(l)\top}_{(k)}$ would have $\prod_{\substack{ i=1 \\ i \neq k}}^{\ord{l}} |\mc{V}_{\phi(l,i)}| \cdot C^{\ord{l} - 1}  = 10^{20}$ entries. %, and tensor $\tst{l}$ would have $\prod_{i=1}^5 |\mc{V}_{\phi(m,i)}| = 10^{20}$ entries.
As a result, computing term (i), $\tst{l}_{(k)}  [\otimes_{i=1}^{\ord{l} \backslash k} \matv{l}{i} ] \tsid^{(l)\top}_{(k)}$, would involve matrix multiplication of a dense $10^{20}$ entry matrix.
However, given the sparsity of $\tst{l}$, one may denote the set of indices of the non-zero entries in tensor $\tnsr{X}$ by $\mathrm{nz}(\tnsr{X}) \coloneqq \{J = (j_{1}, \ldots, j_{N}) \; | \; \tnsr{X}_{j_{1}, \ldots, j_{N}} \neq 0 \}$  and derive the following equivalency
\vspace{-6pt}
\begin{small}
\begin{align*}
& \tst{l}_{(k)}  [\otimes_{i=1}^{\ord{l} \backslash k} \matv{l}{i} ] \tsid^{(l)\top}_{(k)}
= & \sum_{J \in \; \mathrm{nz} (\tst{l})} \tst{l}_{j_{1}, \ldots, j_{o(l)}} \vc{h}(j_k) \prod_{\substack{i=1 \\ i \neq k}}^{\ord{l}} (\matv{l}{i})_{j_i, :},
\end{align*}\vspace{-6pt}
\end{small}\\
where $\prod$ is Hadamard product of a sequence and $\vc{h}(j_k)$ is one-hot column vector of size $|\mc{V}_{\phi(l,k)}|$ that has entry $1$ at index $j_k$.
Computing the right-hand side of this equivalency involves the summation over Hadamard product of a small sequence of small vectors, which has a complexity of $O(\mathrm{nnz} (\tst{l}) \cdot (\ord{l}-1) \cdot C)$ with $\mathrm{nnz} (\cdot)$ being the number of non-zero entries.
In other words, if the previous example comes with $1,000,000$ motif instances, the complexity would decrease from manipulating a $10^{20}$-entry dense matrix to a magnitude of $4 \times 10^{7}$. 

Similarly, by leveraging the sparsity of tensors and composition of dense matrices, one can simplify the computation of term (ii) from multiplication of matrix with $10^{20}$ entries to that with $10^5$ entries; and reduce the calculation of term (iii) from a magnitude of $10^{20}$ to a magnitude of $10^{8}$.
We provide detailed derivation and formulas in the supplementary file.

Finally, we remark that the above computation can be highly parallelized, which has further promoted the efficiency in our implementation.
An empirical efficiency study is available in Section~3\fakeref{} of the supplementary file.
%Likewise, the bottleneck of computing the update rule in Theorem~\ref{thm::update-v} is the computation of the first term on the numerator and the first term on the denominator.
%Fortunately, this problem can also be resolved by exploiting the sparsity of $\tst{m}$ and the composition of $[\otimes_{i=1}^{\ord{l} \backslash k} \matv{l}{i} ] \tsid^{(l)\top}_{(k)}$ in a way similar to Eq.~\eqref{eq::speed-up}, and we omit it due to space limit.
%In practice, the algorithm enjoys an efficiency sublinear to the number of motif instances as we shall present in Section~\ref{sec::exp}.
We summarize the algorithm in Algorithm~\ref{alg::inference}.

%% archive
%Denote $\mathrm{nzv}_{k}(\tnsr{X}) \coloneqq \{(\ldots, j_{k-1}, j_{k+1}, \ldots) \; | \; \text{the fiber } \tnsr{X}_{\ldots, j_{k-1}, j_{k+1}, \ldots} \text{ is non-zero} \}$ the set of indices of the fibers (\ie, vectors) along the $k$-th dimension of the tensor $\tnsr{X}$ that has at least one non-zero entry.
%\begin{align*}
%& \tst{l}_{(k)}  [\otimes_{i=1}^{\ord{l} \backslash k} \matv{l}{i} ] \tsid^{(l)\top}_{(k)}
%= & \sum_{\substack{(\ldots, j_{k-1}, j_{k+1}, \ldots) \\ \in \; \mathrm{nzv}_k (\tst{l})}} \tst{l}_{\ldots, j_{k-1}, :, j_{k+1}, \ldots} \prod_{\substack{i=1 \\ i \neq k}}^{\ord{l}} (\matv{l}{i})_{j_i, :}.
%\end{align*}

\section{Experiments}\label{sec::exp}
We present the quantitative evaluation results on two real-world datasets through multiple tasks and conduct case studies under various circumstances.

\subsection{Datasets and Evaluation Tasks}\label{sec::data}
In this section, we briefly describe (i) the datasets, (ii) the evaluation tasks, and (iii) the metrics used in the experiments.
All of their detailed descriptions are provided in Section~4\fakeref{} of the supplementary file.

\vpara{Datasets.}
We use two real-world HINs for experiments.
%\begin{itemize}
%\item
\textbf{DBLP} is a heterogeneous information network that serves as a bibliography of research in computer science area~\cite{tang2008arnetminer}.
The network consists of 5 types of node: author ($A$), paper ($P$), key term ($T$), venue ($V$) and year ($Y$).
%The key terms are extracted and released by Chen et al.~\cite{chen2017task}. 
%The edge types include authorship, term usage, venue published, year published, and the reference relationship. 
%The first four edge types are undirected, and the last one is directed. 
%The schema of the DBLP network is shown in Figure~\ref{fig::dblp-schema}.
In DBLP, we select two candidate motifs for all applicable methods, including $AP4TPA$ and $APPA$. %, where $APPA$ is a also a meta-path representing author writes a paper that refers another paper written by another author. % and $AP4TPA$ was introduced in Section~\ref{sec::motivation}.
%\item
\textbf{YAGO} is a knowledge graph constructed by merging Wikipedia, GeoNames and WordNet. YAGO dataset consists of 7 types of nodes: person (\textit{P}), organization (\textit{O}), location (\textit{L}), prize (\textit{R}), work (\textit{W}), position (\textit{S}) and event (\textit{E}). 
%There are 24 types of edges in the network, with 19 undirected edge types and 5 directed edge types as shown by the schema of the YAGO network in Figure~\ref{fig::yago-schema}.
%Its schema is provided in the supplementary file.
In YAGO, the candidate motifs used by all compared methods include $P^{6}O^{23}L$, $P^{7}O^{23}L$, $P^{8}O^{23}L$, $2P2W$, $3PW$. %, where the first three are also meta-paths.
%$2P2W$ is the motif that 2 people simultaneously co-created two pieces of work, and $3PW$ is the motif that 3 people who created, directed, and acted in a piece of work, respectively.
%\end{itemize}
%
%
%\begin{figure}[t]
%  \centering
%  \includegraphics[width=.8\linewidth]{figures/yago_schema_pkdd}
%    %\vspace{-9pt}
%    \caption{The schema of YAGO~\cite{shi2018easing}.}\label{fig::yago-schema}
%    %\vspace{-3pt}
%\end{figure}

\vpara{Evaluation tasks.}
In order to evaluate models' capability in reflecting different user guidance, we use two sets of labels on authors to conduct two tasks in DBLP similar to previous study~\cite{sun2012integrating}.
Additionally, we design another task on YAGO with labels on persons.
%We will release datasets and labels used in the experiment once the paper is published.
%\begin{itemize}
%\item
\textbf{DBLP-group} -- 
Clustering authors to $5$ research groups where they graduated. %, which is an expanded label set from the ``four-group dataset''~\cite{sun2012integrating}. 
%The ``four-group dataset'' includes researchers from four renowned research groups led by Christos Faloutsos, Michael I. Jordan, Jiawei Han, and Dan Roth. 
%Additionally, we add another group of researchers, who have collaborated with at least one of the researchers in the ``four-group dataset'' and label them as the fifth group with the intention to involve more subtle semantics in the original HIN.
$5\%$ of the $250$ authors with labels are randomly selected as seeds from user guidance.
%We did not use $1\%$ for seed ratio as in the following two tasks because the number of authors to be clustered in this task is small.
%The resulted HIN processed as such consists of 19,500 nodes and 108,500 edges.
%\item
\textbf{DBLP-area} -- 
Clustering authors to $14$ research areas. %, which is expanded from the ``four-area dataset''~\cite{sun2012integrating}, where the definition of the 14 areas is derived from the Wikipedia page: List of computer science conferences\footnote{https://en.wikipedia.org/wiki/List   of   computer   science   conferences}. 
$1\%$ of the $7,165$ authors with labels are randomly selected as seeds from user guidance.
%The HIN processed in this way has 16,100 nodes and 30,239 edges.
%\item
\textbf{YAGO} -- 
Clustering people to 10 popular countries in the YAGO dataset. 
%We knock out all edges with edge type wasBornIn, and if a person had an edge with one of the 10 countries, we assign this country to be the label of this person.
%Additionally, to avoid making our task trivial, we remove all other types of edges between person and location. 
$1\%$ of the $11,368$ people are randomly selected as seeds from user guidance.
%There are 17,109 nodes and 70,251 edges in the processed HIN.
%\end{itemize}

\vpara{Evaluation metrics.}
We use three metrics to evaluate the quality of the clustering results generated by each model: Accuracy (Micro-F1), Macro-F1, and NMI.
%\textbf{Accuracy} refers to a measure of statistical bias. More precisely it is defined by the division of the number of correctly labeled data by the total size of the dataset. 
Note that in multi-class classification tasks, accuracy is always identical to Micro-F1.
%\textbf{Macro-F1} refers to the arithmetic mean of the F1 score across all different labels in the dataset, where the F1 score is the harmonic mean of precision and recall for a specific label. 
%\textbf{NMI} is the abbreviation for normalized mutual information. Numerically, it is defined as the division of mutual information by the arithmetic mean of the entropy of each label in the data.
For all these metrics, higher values indicate better performance.

\subsection{Baselines and Experiment Setups}
\vpara{Baselines.} We use five different baselines to obtain insight on different aspects of the performance of \mochin.
%\begin{itemize}
%\item
\textbf{KNN} is a classification algorithm that assigns the label of each object in the test set is according to its nearest neighbors.
%This is a homogeneous method that does not distinguish different node types. 
In our scenario, the distance between two nodes is defined as the length of the shortest path between them.
%\item
\textbf{KNN+Motifs} 
%serves as a direct comparison to the proposed \mochin model, since KNN+Motifs can also 
uses signals generated by motifs, but does not directly model all players in higher-order interactions.
To extract information from motifs, we construct a motif-based network for each candidate motif, where an edge is constructed if two nodes are matched to a motif instance in the original HIN.
KNN is then applied to each motif-based network.
Finally, a linear combination is applied to the outcome probability matrices generated by KNN from the motif-based networks and the original HIN with weights tuned to the best.
%When using this baseline method, we additionally add $APVPA$ into the set of candidate motifs for both DBLP tasks and add $P^{14}O^{14}P$, $P^{15} O^{15}P$, and $P^{16}O^{16}P$ for the YAGO task.
%\item
\textbf{GNetMine}~\cite{ji2010graph}
is a graph-based regularization framework to address the transductive classification problem in HINs. 
This method only leverages edge-level information without considering structural patterns such as meta-paths or motifs.
%\item
\textbf{PathSelClus}~\cite{sun2012integrating}
is a probabilistic graphical model that performs clustering tasks on HINs by integrating meta-path selection with user-guided clustering.
For this baseline, we additionally add $APVPA$, $APTPA$, $APT$, $APA$, and, $APAPA$ into the set of candidate meta-paths for both DBLP tasks as suggested by the original paper~\cite{sun2012integrating} and add $P^{14}O^{14}P$, $P^{15} O^{15}P$, and $P^{16}O^{16}P$ for YAGO task.
%\item
\textbf{TGS}~\cite{carranza2018higher} leverages motifs but does not directly model each occurrence of higher-order interaction.
It is hence another direct comparison to \mochin{}, besides KNN+Motifs, which is used to analyze the utility of comprehensively transcribing motif instances into tensors.
As the authors did not discuss how to inject user guidance into their basic bipartitioning clustering algorithm, we apply multi-class logistic regression on the accompanied typed-graphlet spectral embedding algorithm proposed in the same paper.
The typed-graphlet adjacency matrices of multiple motifs are summed together to derive the input for the algorithm as the author suggested in the paper.
%\end{itemize}

\vpara{Experiment setups.}
For \mochin, we set hyperparameters $\theta = 1$, $\rho = 100$ and $\lambda = 0.0001$ across all tasks in our experiments. 
For each model involving motifs, edge-level motifs corresponding to the edge types are included into the set of candidate motifs.
For each baseline in each task, we always tune its hyperparameters to achieve the best performance.

\subsection{Quantitative Evaluation Result}

\begin{table*}[t]
\centering
\caption{Quantitative evaluation on clustering results in three tasks.}\label{tab::quant-eval}
%\vspace{-9pt}
\resizebox{\textwidth}{!}{
\begin{tabular}{l | c | c | c | c | c | c | c | c | c}
\toprule \hline
Task &  \multicolumn{3}{c|}{DBLP-group}  & \multicolumn{3}{c|}{DBLP-area}  & \multicolumn{3}{c}{YAGO} \\ \hline
Metric & Acc./Micro-F1 & Macro-F1 & NMI & Acc./Micro-F1 & Macro-F1 & NMI & Acc./Micro-F1 & Macro-F1 & NMI \\ \hline \hline
KNN & 0.4249 & 0.2566 & 0.1254 & 0.4107 & 0.4167 & 0.2537 & 0.3268 & 0.0921 & 0.0810 \\ \hline
KNN+Motifs & 0.4549 & 0.2769 & 0.1527 & 0.4811 & 0.4905 & 0.3296 & 0.3951 & 0.1885 & 0.1660  \\ \hline
GNetMine~\cite{ji2010graph} & 0.5880 & 0.6122 & 0.3325 & 0.4847 & 0.4881 & 0.3469 & 0.3832 & 0.2879 & 0.1772  \\ \hline 
PathSelClus~\cite{sun2012integrating} & 0.5622 & 0.5535 & 0.3246 & 0.4361 & 0.4520 & 0.3967 & 0.3856 & 0.3405 & 0.2864 \\ \hline
TGS~\cite{carranza2018higher}  & 0.6609 & 0.6513 & 0.3958 & 0.4391 & 0.4365 & 0.2790 & 0.6058 & 0.3564 & 0.4406 \\ \hline \hline
MoCHIN & \textbf{0.7382} & \textbf{0.7387} & \textbf{0.5797} & \textbf{0.5318} & \textbf{0.5464}& \textbf{0.4396} & \textbf{0.6134}  & \textbf{0.5563} & \textbf{0.4607}  \\ \hline
\bottomrule
\end{tabular}
}
%\vspace{6pt}
\end{table*}

%We quantitatively evaluate the effectiveness of the proposed \mochin model against the baselines and 
We report the main quantitative results in Table~\ref{tab::quant-eval}.
Overall, \mochin uniformly outperformed all baselines in all three tasks under all metrics.
Note that these three metrics measure different aspects of the model performance. 
For instance, in the DBLP-area task, PathSelClus outperforms GNetMine under Macro-F1 and NMI, while GNetMine outperforms PathSelClus under Acc./Micro-F1.
Achieving superior performance uniformly under all metrics is hence strong evidence that \mochin with higher-order interaction directly modeled is armed with greater modeling capability in the task of user-guided HIN clustering.

\vpara{\mochin prevails in user-guided clustering by exploiting signals from motifs more comprehensively.}
Recall that KNN+Motifs, TGS, and \mochin all exploit signals from motifs.
However, the two baselines do not directly model each occurrence of motif instances and only preserve pairwise or edge-level information.
In our experiments, even though TGS can generally outperform other baselines, it alongside KNN+Motifs still cannot generate results as good as \mochin, which demonstrates the utility of more comprehensively exploiting signals from motifs as \mochin does.
%In fact, the set of non--edge-level motifs used in the baseline is always a superset of that used in \mochin.
%We interpret this result as follows: although \mochin uses fewer motifs, by modeling all players in the higher order interaction, it implicitly captures information carried by other motifs, which justifies the use of a more complex model.
We interpret this result as when user guidance is limited, a fine-grained understanding of the rich semantics of an HIN is instrumental in dissecting users' intention and generating desirable results.

%\vpara{Heterogeneous methods outperform homogeneous method.}
%Among all $5$ methods compared in the experiments, KNN does not use any type information, while the other four consider heterogeneity in the network by distinguishing different node types and/or leveraging HIN structure patterns such as meta-paths and motifs.

\hide{
\vpara{KNN is disadvantaged on imbalanced data when the supervision is weak.}
\cmr{to keep this or not? if keep, use link it to why \mochin is good}
Across all the three tasks conducted in the experiments, the ground truth labels in the YAGO dataset are the most imbalanced.
As presented in Table~\ref{tab::quant-eval}, KNN performs notably worse on YAGO with $1\%$ seed ratio under Macro-F1 and NMI, which are more sensitive to model performance on rare classes compared to Accuracy (Micro-F1).
In other words, KNN tends to achieve inferior results on rare classes when supervision is weak, and data is imbalanced.
We recommend using other heterogeneous methods that consider the type information in this scenario.
The results in Section~\ref{sec::varied-seed-ratio} can further validate this point, where experiments with varied seed ratio are presented.
}

\begin{table}[t]
\centering
\caption{Ablation study of the MoCHIN model on the DBLP-group task with the non--edge-level motifs, $APPA$ and $AP4TPA$, optionally removing from the full model.}\label{tab::ablation}%\vspace{-12pt}
\begin{tabular}{ c c c c c}  
    \resizebox{.48\textwidth}{!}{
    \begin{tabular}{l | c | c | c | c}
    \multicolumn{5}{}{} \\
    \toprule \hline
    \multirow{ 2}{* }{Metric} & \multirow{ 2}{* }{Acc./Micro-F1} & \multirow{ 2}{* }{Macro-F1} & \multirow{ 2}{* }{NMI} & Result for \\ 
    & & & & Eric Xing \\ \hline \hline
    W/o both & 0.6567 & 0.6411 & 0.5157 & \xmark  \\ \hline
    W/  APPA & 0.7039 & 0.7062 & 0.5166 & \xmark \\ \hline
    W/  AP4TPA & 0.6781 & 0.6589 & 0.5502 & \cmark \\ \hline 
    Full model &  \textbf{0.7382} & \textbf{0.7387} & \textbf{0.5797}   & \cmark \\ \hline
    \bottomrule
    \end{tabular}
    }
    &
    &
    \begin{minipage}{.13\textwidth}
      \includegraphics[width=\linewidth]{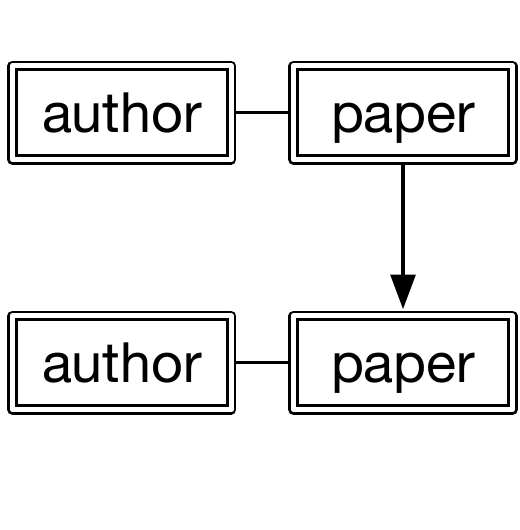}
    \end{minipage}
    &
    &
    \begin{minipage}{.22\textwidth}
      \includegraphics[width=\linewidth]{figures/ap4tpa}
    \end{minipage} %\vspace{-3pt}
    \\
   & & { \small The APPA motif.} & & {\small The AP4TPA motif.} 
\end{tabular}

%\vspace{-12pt}
\end{table}

%\subsection{Impact of Candidate Motif Choice}
\vpara{Impact of candidate motif choice.}
In this section, we study how the choice of candidate motifs impacts \mochin and additionally use the concrete example in Figure~\ref{fig::subnetwork} to understand the model outputs.
Particularly, we conducted an ablation study by taking out either or both of the two non--edge-level motifs, $APPA$ and $AP4TPA$, in the DBLP-group task and reported the result in Table~\ref{tab::ablation}. 
%As introduced in Section~\ref{sec::data}, the candidate motifs \mochin used in the DBLP tasks are all edge types with two non--edge-level motifs: $APPA$ and $AP4TPA$.
%We conducted an ablation study by taking out either or both of the two non--edge-level motifs and compared with the original full model in the DBLP-group task, and the results are reported in Table~\ref{tab::ablation}. 
The full \mochin model outperformed all partial models, demonstrating the utility of these motifs in clustering.

Moreover, we scrutinized the concrete example in Figure~\ref{fig::subnetwork} and checked how each model assigned cluster membership for Eric Xing.
The result is also included in Table~\ref{tab::ablation}, which shows only the model variants with $AP4TPA$ made the correct assignment on Eric Xing.
In Section~2\fakeref{} of the supplementary file, a visualization of this ablation study is provided to further corroborate our observation.

\hide{

\begin{figure}[ht]
  \centering
  \begin{subfigure}[t]{.4\linewidth}
    \centering\includegraphics[width=\linewidth]{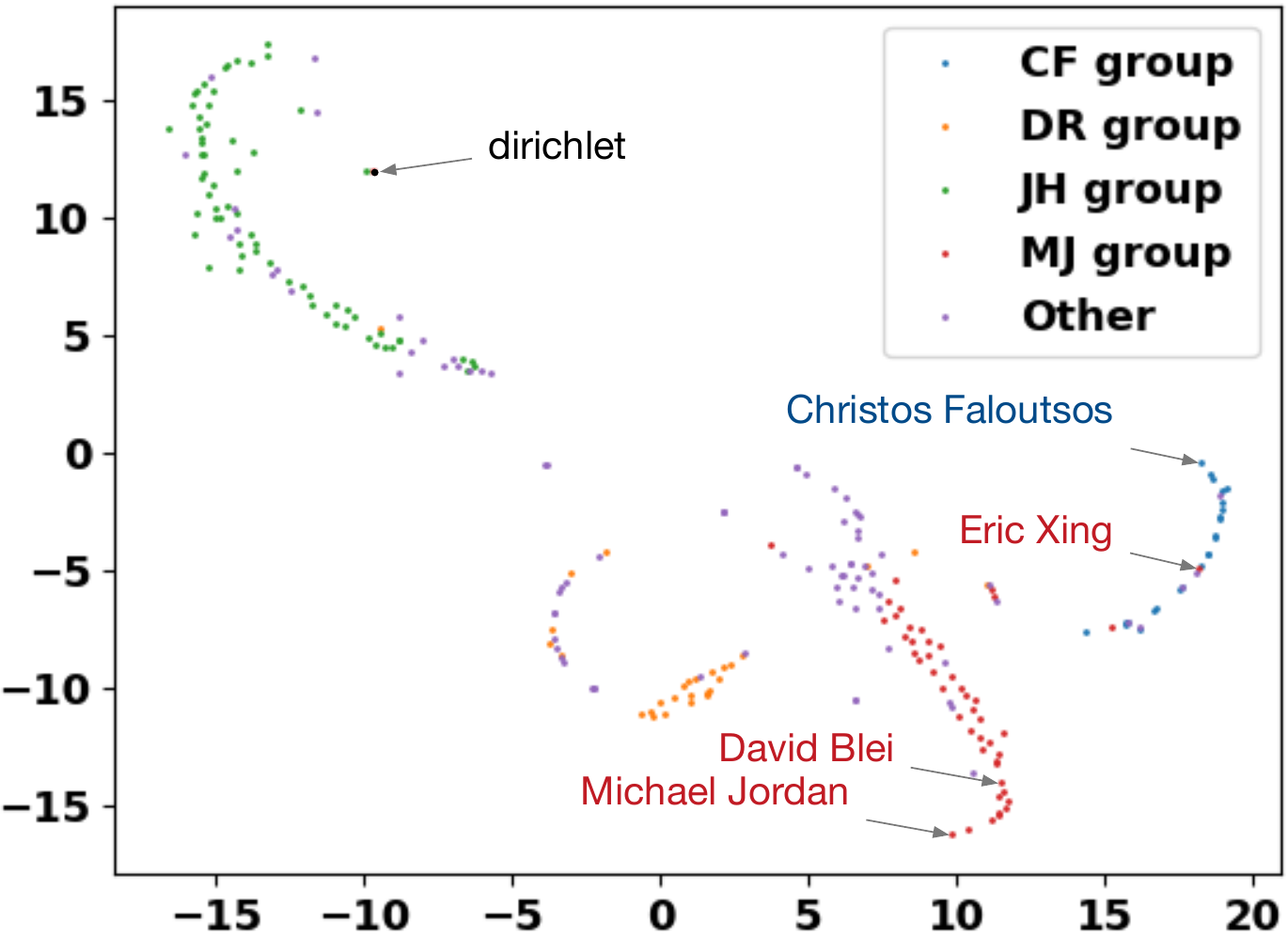}
    \caption{W/o both.}
  \end{subfigure}
  \qquad \qquad
  \begin{subfigure}[t]{.4\linewidth}
    \centering\includegraphics[width=\linewidth]{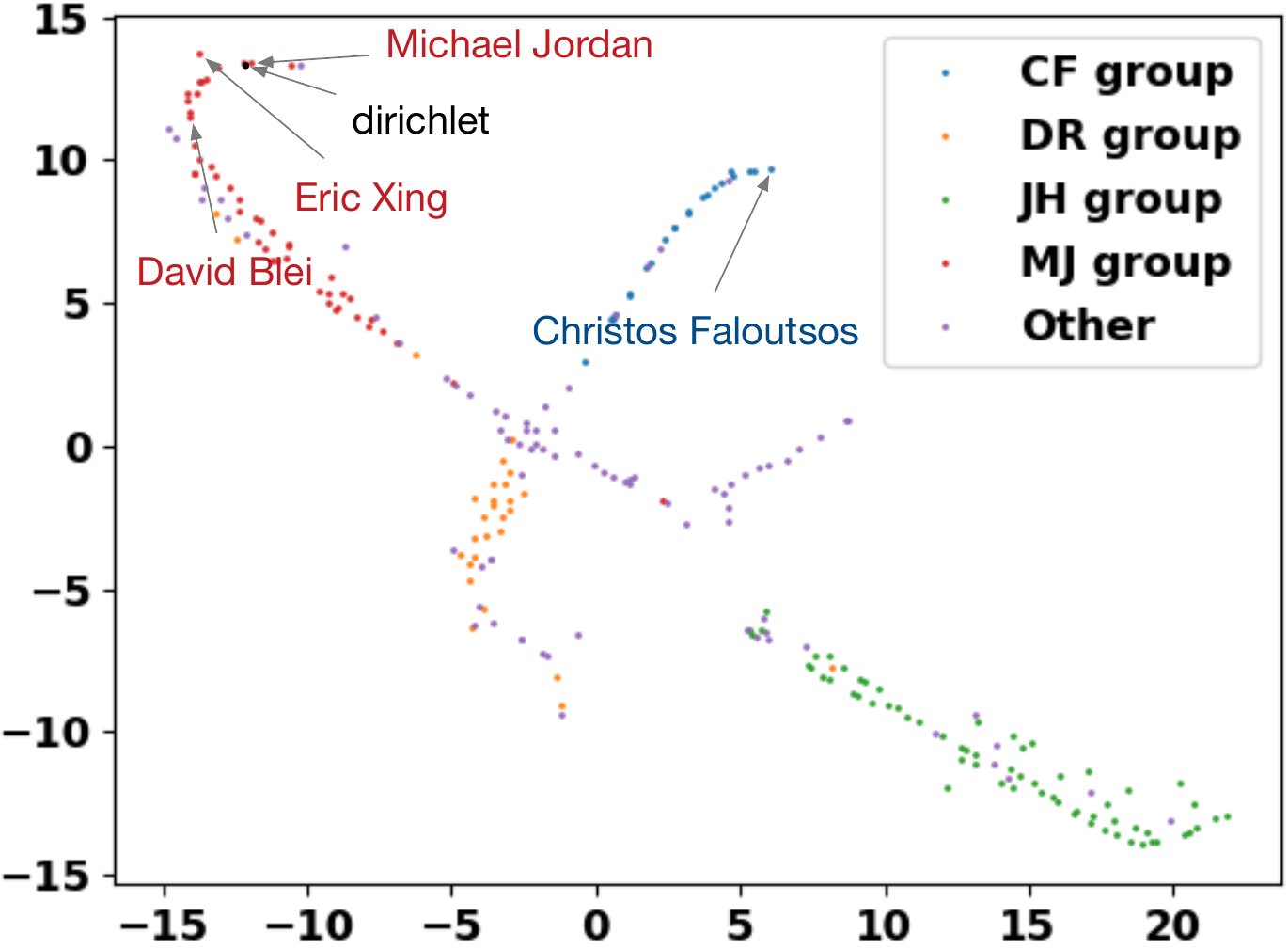}
    \caption{Full model}
  \end{subfigure}%\vspace{-6pt}
  \caption{Visualization of the ablation study where the author nodes are color-coded according to the truth label.}\label{fig::vis}
\end{figure}

To better understand this result, we further visualized the inferred membership of each node in Figure~\ref{fig::vis} by projecting its corresponding column in the consensus matrix $\matvcs{t}$ using t-Distributed Stochastic Neighbor Embedding (t-SNE).
As discussed in Section~\ref{sec::motivation}, \textit{dirichlet} reflects a distinctive facet of the relationship between \textit{Xing} and \textit{Blei} pertaining to their graduating group.
The full model containing $AP4TPA$ inferred all of them to be close under the user guidance concerning research group.
In contrast, the partial model with only edge-level motifs not only mistakenly assigned \textit{Xing} to \textit{Faloutsos}'s group but also learned \textit{dirichlet} to be far away from either Xing or Blei.
This observation echos the intuition discussed in Section~\ref{sec::motivation} that modeling higher-order interaction can introduce a richer pool of signals, and such modeling should be comprehensive and fine-grained in the task of user-guided clustering.
}

\subsection{Varied Seed Ratio}\label{sec::varied-seed-ratio}
\begin{figure}[t]
  \centering
  \begin{subfigure}[t]{.48\linewidth}
    \centering\includegraphics[width=\linewidth]{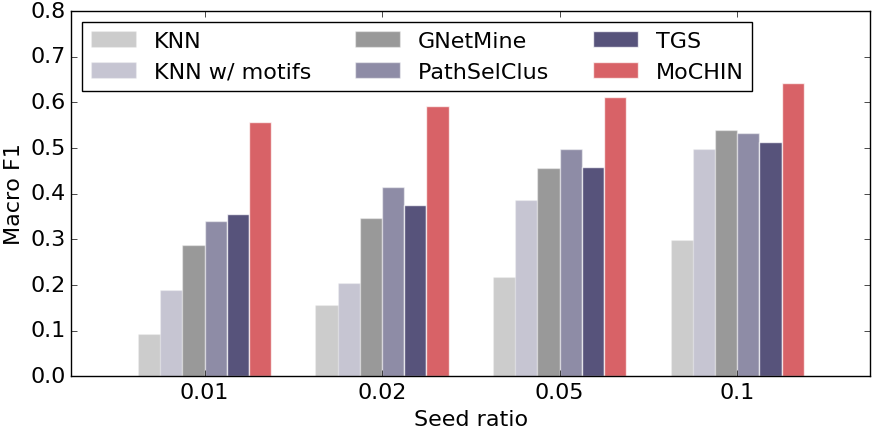}
    \caption{Macro-F1.}
  \end{subfigure}
  \begin{subfigure}[t]{.48\linewidth}
    \centering\includegraphics[width=\linewidth]{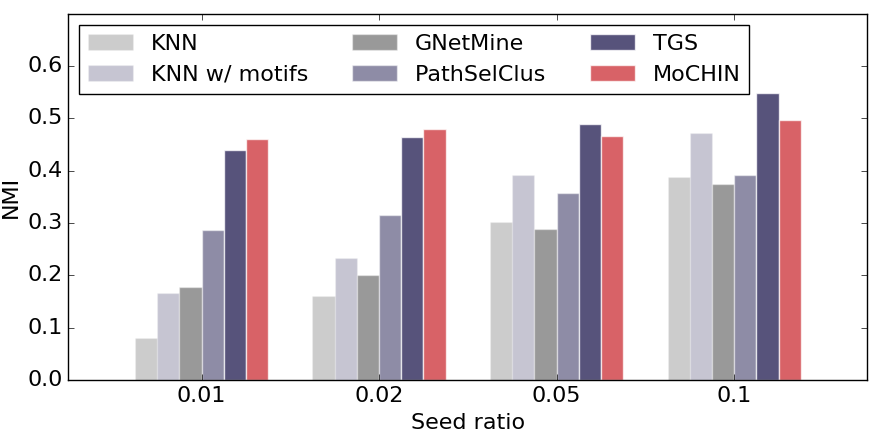}
    \caption{NMI.}
  \end{subfigure}\vspace{-6pt}
  \caption{Quantitative evaluation on the YAGO task under varied seed ratio. }\label{fig::varying-seed-ratio}
  \vspace{-12pt}
\end{figure}

In addition to using $1\%$ people as seeds for the YAGO task reported in Table~\ref{tab::quant-eval}, we experiment under varied seed ratio $2\%$,  $5\%$,  and $10\%$.
The results are reported in Figure~\ref{fig::varying-seed-ratio}.
We omit Accuracy (Micro-F1), which has a similar trend with NMI.

For all methods, the performance increased as the seed ratio increased. %, which was a natural outcome of progressively stronger supervision.
Notably, \mochin outperformed most baselines, especially when seed ratio is small.
%Notably, the difference in performance between \mochin and the baselines shrank as seed ratio increased.
%This suggests when supervision is strong enough, the pairwise edge level signal can provide progressively sufficient information to obtain reasonable results.
%On the other hand, \mochin is particularly attractive when supervision is weak for being able to extract more subtle information from limited data.
This suggests \mochin is particularly useful when users provide less guidance for being able to better exploit subtle information from limited data.
Note that higher seed ratio is uncommon in practice since it is demanding for users to provide more than a few seeds.

Lastly, an efficiency study that empirically evaluates the proposed algorithm is provided in Section~3\fakeref{} of the supplementary file.

\hide{
Probably hide Figure~\ref{fig::instance}, since it does not support our key claim on comprehensively model higher-order interaction and may fire back if people say with a motif, nothing in the middle is needed.
\begin{figure}[t]
  \centering
  \begin{subfigure}[t]{0.46\linewidth}
    \centering\includegraphics[width=\linewidth]{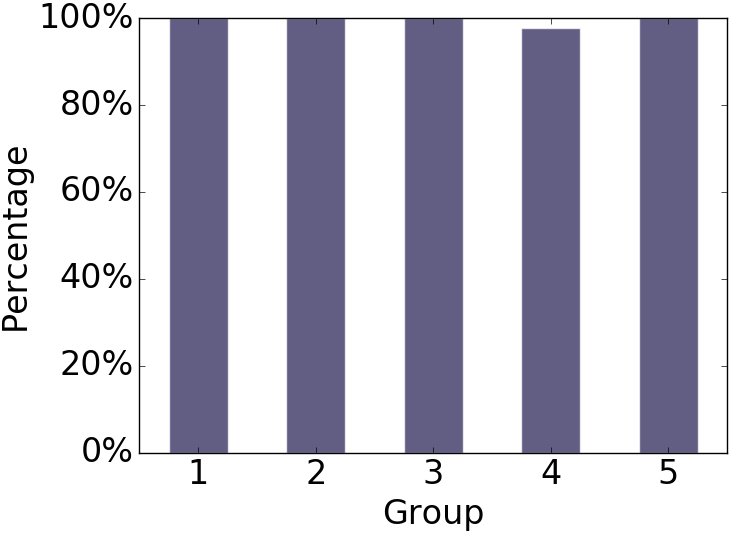}
    \caption{Percentage of authors matched by metapath APTPA.}\label{fig::mp-ratio}
  \end{subfigure}
  \hspace{6pt}
  \begin{subfigure}[t]{0.46\linewidth}
    \centering\includegraphics[width=\linewidth]{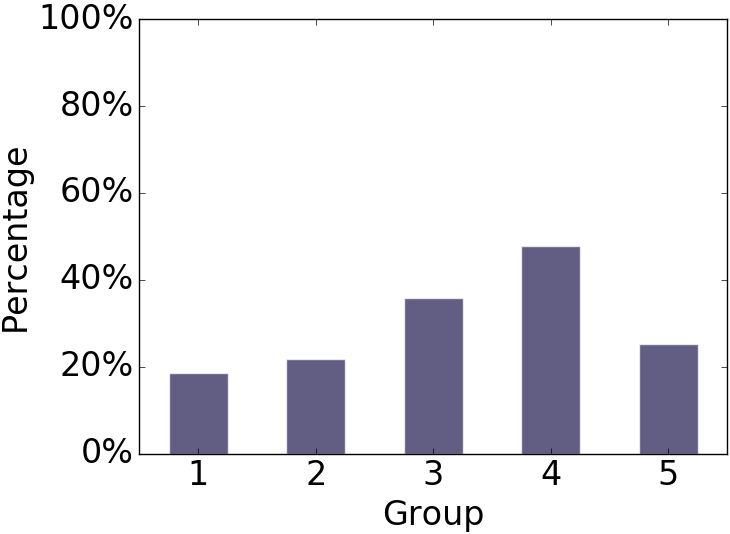}
    \caption{Percentage of authors matched by motif AP4TPA.}\label{fig::mt-ratio}
  \end{subfigure}
  \hspace{6pt}
    \begin{subfigure}[t]{0.46\linewidth}
    \centering\includegraphics[width=\linewidth]{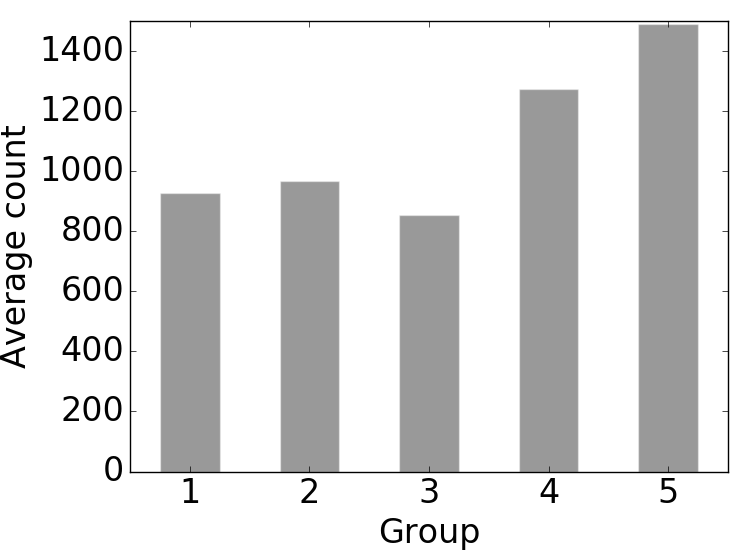}
    \caption{Average count of metapath APTPA instances.}\label{fig::mp-count}
  \end{subfigure}
  \hspace{9pt}
  \begin{subfigure}[t]{0.46\linewidth}
    \centering\includegraphics[width=\linewidth]{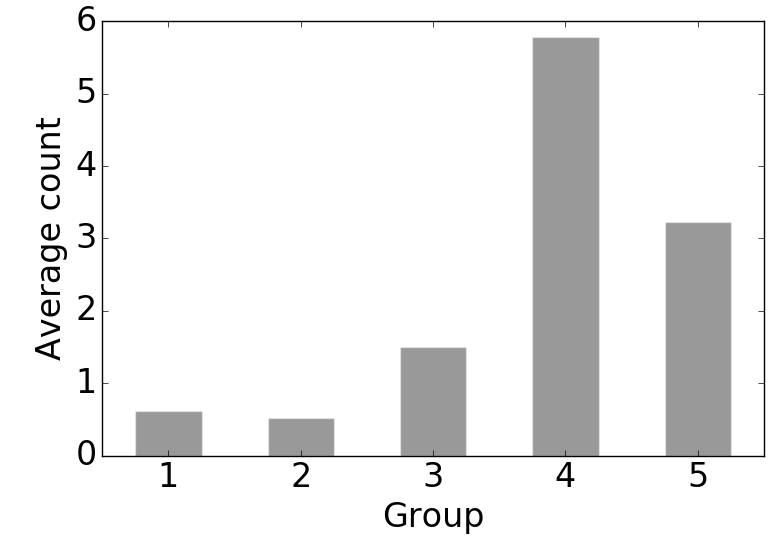}
    \caption{Average count of motif AP4TPA instances}\label{fig::mt-count}
  \end{subfigure}
  \hspace{3pt}
  \caption{Instances involving Eric Xing are found under metapath APTPA and motif AP4TPA. In Figure~\ref{fig::mp-ratio} and \ref{fig::mt-ratio}, the percentage of authors linked to Eric Xing from each group is presented. Figure~\ref{fig::mp-count} and \ref{fig::mt-count} give the average count of metapath or motif instances covering both Eric Xing and one author from each group.}\label{fig::instance}
\end{figure}

}

%!TEX root = hin_motif_clus.tex

\section{Discussion, Conclusion, and Future Work}
One limitation of \mochin{} is that it may not be easily applied to very large datasets even with speed-up methods due to the complexity of the model itself.
However, \mochin would stand out in the scenario where fine-grained understanding of the network semantics is needed.
In the experiment, we have shown that \mochin can scale to HINs with tens of thousands of nodes.
We note that for user-guided clustering, it is possible the users are mostly interested in the data instances most relevant to their intention, which could be a subset of a larger dataset.
For instance, if a data mining researcher wanted to cluster DBLP authors by research group, it is possible they would not care about the nodes not relevant to data mining research.
As such the majority of the millions of nodes in DBLP can be filtered out in preprocessing, and this user-guided clustering problem would become not only manageable to \mochin but also favorable due to \mochin{}'s capability in handling fine-grained semantics.
Moreover, in the case where the network is inevitably large, one may trade the performance of \mochin for efficiency by using only relatively simple motifs.
It is also worth noting that incremental learning is possible for \mochin{} -- when new nodes are available, one do not have to retrain the model from scratch.

In conclusion, we studied the problem of user-guided clustering in HINs with the intention to model higher-order interactions.
We identified the importance of modeling higher-order interactions without collapsing them into pairwise interactions and proposed the \mochin algorithm. 
Experiments validated the effectiveness of the proposed model and the utility of comprehensively modeling higher-order interactions. 
Future works include exploring further methodologies to join signals from multiple motifs, which is currently realized by a simple linear combination in the \mochin model.
Furthermore, as the current model takes user guidance by injecting labels of the seeds, it is also of interest to extend \mochin to the scenario where guidance is made available by must-link and cannot-link constraints on node pairs.

\vpara{Acknowledgments.}
This work was sponsored in part by U.S. Army Research Lab. under Cooperative Agreement No. W911NF-09-2-0053 (NSCTA), DARPA under Agreement No. W911NF-17-C-0099, National Science Foundation IIS 16-18481, IIS 17-04532, and IIS-17-41317, DTRA HDTRA11810026, and grant 1U54GM114838 awarded by NIGMS through funds provided by the trans-NIH Big Data to Knowledge (BD2K) initiative (www.bd2k.nih.gov). Any opinions, findings, and conclusions or recommendations expressed in this document are those of the author(s) and should not be interpreted as the views of any U.S. Government. The U.S. Government is authorized to reproduce and distribute reprints for Government purposes notwithstanding any copyright notation hereon.

\bibliographystyle{splncs04}
%\balance
\bibliography{sigproc} 

\newpage

\chapter*{Supplementary Materials}

\setcounter{section}{0}

\section{Additional Proof and Formulas}
\subsection{The Proof for Theorem~1\fakeref{} in the Main File}
Inspired by prior art on non-negative matrix factorization~\cite{lee2001algorithms}, we provide the proof for Theorem~1 in the main file on tensor factorization as follows.

\begin{proof}
With the equivalency given by Lemma~1\fakeref{} in the main file
\begin{small}
\begin{align*}
\Big\lVert\tst{m} - \tsid^{(m)} & \times_{i=1}^{\ord{m}} \matv{m}{i} \Big\rVert_{F} = \norm{ \tst{m}_{(k)} - \matv{m}{k} \tsid^{(m)}_{(k)} [\otimes_{i=1}^{\ord{m} \backslash k} \matv{m}{k} ]\trans  }_F,
\end{align*}
\end{small}
we construct the auxiliary function
\begin{align*}
 \mc{Z}(\matv{l}{k}, \tmatv) 
= & \sum_{s,t} \big\{ 
\subst{\matv{l}{k} \tsid^{(l)}_{(k)} [\otimes_{i=1}^{N \backslash k} \matv{l}{i} ]\trans [\otimes_{i=1}^{N \backslash k} \matv{l}{i} ] \tsid^{(l)\top}_{(k)}} \subst{\tmatv}^2 / \subst{\matv{l}{k}} \\
- & 2 \subst{\tst{l}_{(k)}  [\otimes_{i=1}^{N \backslash k} \matv{l}{i} ] \tsid^{(l)\top}_{(k)}} \subst{\matv{l}{k}} (1 + \log \frac{\subst{\tmatv}}{\subst{\matv{l}{k}}}) \\
+ & \theta (1 - \eta^{l}_{k})^2 \subst{\tmatv}^2 
- 2 \theta (1 - \eta^{l}_{k}) (\matvcs{\phi(l, k)} - \eta^{l}_{k} \matv{l}{k}) \\
\cdot & \subst{\matv{l}{k}} (1 + \log \frac{\subst{\tmatv}}{\subst{\matv{l}{k}}})
+ \theta \sum_{\phi(m,i) = \phi(l,k) }^{ (m,i) \neq (l,k)} \eta^{l2}_{k} \subst{\tmatv}^2 \\
- & 2 \subst{[\theta \sum_{\phi(m,i) = \phi(l,k) }^{ (m,i) \neq (l,k)} \eta^{l}_{k} (\matv{m}{i} - \matvcs{\phi(l, k)} + \eta^{l}_{k} \matv{l}{k})]^{+}} \\ 
\cdot & \subst{\matv{l}{k}} (1 + \log \frac{\subst{\tmatv}}{\subst{\matv{l}{k}}})
+ \subst{[\theta \sum_{\phi(m,i) = \phi(l,k) }^{ (m,i) \neq (l,k)} \eta^{l}_{k} (\matv{m}{i} \\ 
- & \matvcs{\phi(l, k)} + \eta^{l}_{k} \matv{l}{k})]^{-}}  (\subst{\matv{l}{k}}^2 + \subst{\tmatv}^2) / \subst{\matv{l}{k}} \\
+ & \rho \subst{\mat{M}^{\phi(l,k)} \circ (\matvcs{\phi(l,k)} - \eta^{l}_{k} \matv{l}{k} + \eta^{l}_{k} \tmatv )}^2 \\
+ & \lambda (\subst{\matv{l}{k}}^2 + \subst{\tmatv}^2) / (2 \subst{\matv{l}{k}})
\big\}.
\end{align*}
Straightforward derivation can show the following three relations hold: 
\begin{enumerate}
\item
$\mc{Z}(\matv{l}{k}, \matv{l}{k}) = \mc{O}(\matv{l}{k})$,
\item
$\mc{Z}(\matv{l}{k}, \tmatv) \geq \mc{O}(\tmatv)$, and 
\item
$\mc{Z}(\matv{l}{k}, \tmatv)$ is convex with respect to $\tmatv$.
\end{enumerate}
%(i) $\mc{Z}(\matv{l}{k}, \matv{l}{k}) = \mc{O}(\matv{l}{k})$, (ii) $\mc{Z}(\matv{l}{k}, \tmatv) \geq \mc{O}(\tmatv)$, and (iii) $\mc{Z}(\matv{l}{k}, \tmatv)$ is convex with respect to $\tmatv$.
Therefore, by setting $\frac{\partial}{\partial \tmatv} \mc{Z}(\matv{l}{k}, \tmatv) = 0$, one can find $\mc{Z}(\matv{l}{k}, \tmatv)$ is minimized at $\tmatv = \tmatv_{\mathrm{opt}}$, where $\tmatv_{\mathrm{opt}}$ is the righthand side of Eq.~(5)\fakeref{} in the main file, and 
$\mc{O}(\matv{l}{k}) = \mc{Z}(\matv{l}{k}, \matv{l}{k}) \geq \mc{Z}(\matv{l}{k}, \tmatv_{\mathrm{opt}}) \geq \mc{O}(\tmatv_{\mathrm{opt}})$.
It follows that setting $\matv{l}{k}$ to $\tmatv_{\mathrm{opt}}$ monotonically decreases the objective function $\mc{O}$ which is exactly the update rule in Theorem~1.
\end{proof}

\subsection{Omitted Formulas for Inference Speed-Up Methods}
The first term in the denominator of Eq.~(5)\fakeref{} in the main file, $\matv{l}{k} \tsid^{(l)}_{(k)} [\otimes_{i=1}^{\ord{l} \backslash k} \matv{l}{i} ]\trans $ $ [\otimes_{i=1}^{\ord{l} \backslash k} \matv{l}{i} ] \tsid^{(l)\top}_{(k)}$, again involves matrix multiplication of the huge dense matrix $[\otimes_{i=1}^{\ord{l} \backslash k} \matv{l}{i} ] \tsid^{(l)\top}_{(k)}$.
Leveraging the composition of $[\otimes_{i=1}^{\ord{l} \backslash k} \matv{l}{i} ] \tsid^{(l)\top}_{(k)}$, one can show that
$$
\matv{l}{k} \tsid^{(l)}_{(k)} [\otimes_{i=1}^{\ord{l} \backslash k} \matv{l}{i} ]\trans [\otimes_{i=1}^{\ord{l} \backslash k} \matv{l}{i} ] \tsid^{(l)\top}_{(k)} = \matv{l}{k} \prod_{\substack{i=1 \\ i \neq k}}^{\ord{l}} \left(\matv{l}{i}{}\trans \matv{l}{i}\right).
$$
where $\prod$ is Hadamard product of a sequence.
As such, instead of multiplying a huge dense matrix, one may only compute Hadamard product and matrix multiplication over a few relatively small matrices.
Note that in the example provided in Section~6\fakeref{} of the main file, $[\otimes_{i=1}^{\ord{l} \backslash k} \matv{l}{i} ] \tsid^{(l)\top}_{(k)}$ has $10^{20}$ entries, while $\matv{l}{i}$ has only $10000 \times 10 = 10^{5}$ entries and $\ord{l} = 5$.

Lastly, evaluating the loss function Eq.~(3)\fakeref{} in the main file for determining convergence involves the computation of the Frobenius norm of its first term, \ie, $\tst{m} - \tsid^{(m)}\times_{i=1}^{\ord{m}} \matv{m}{i}$, which is a huge, dense tensor.
%Consider the scenario that an HIN motif involved $5$ nodes and each node type had $10,000$ node instances.
%Then this dense tensor would have $\prod_{i=1}^5 |\mc{V}_{\phi(m,i)}| = 10000^5 = 10^{20}$ entries, which is computationally forbidding to traverse and compute its Frobenius norm.
%To overcome this challenge, we develop a speed-up trick by leveraging the fact that out of all consistent combinations of nodes, only a small portion can form motif instances in real-world networks, and this fact guarantees $\tst{m}$ to be a sparse tensor.
Again by exploiting the desirable sparsity property of $\tst{m}$, we can calculate the Frobenius norm of $\tst{m} - \tsid^{(m)}\times_{i=1}^{\ord{m}} \matv{m}{i}$ as follows
%\begin{small}
%\vspace{-6pt}
\begin{align*} 
&\norm{\tst{m} - \tsid^{(m)}\times_{i=1}^{\ord{m}} \matv{m}{i}}_{F}^2 \\
= & \norm{\tst{m}}_{F}^2 - 2 \norm{\tst{m} \circ \tsid^{(m)}\times_{i=1}^{\ord{m}} \matv{m}{i}}_1  + \norm{\tsid^{(m)}\times_{i=1}^{\ord{m}} \matv{m}{i}}_{F}^2 \\
= & \norm{\tst{m}}_{F}^2 - 2 \sum_{j_1, \ldots, j_\ord{m}} ( \tst{m} )_{j_1, \ldots, j_{\ord{m}}}  \sum_{c=1}^C \prod_{i=1}^{\ord{m}} (\matv{m}{i})_{j_i, c} \\
+ & \sum_{c_1 = 1}^C \sum_{c_2 = 1}^C \prod_{i=1}^{\ord{m}} (\matv{m}{i})_{:, c_1}\trans (\matv{m}{i})_{:, c_2}.  \numberthis\label{eq::speed-up-obj}
\end{align*}
%\end{small}
This equivalency transforms the computation of a dense and potentially high-order tensor to that of a sparse tensor accompanied by a couple of matrix manipulation.
%Denote $\mathrm{nnz}(\cdot)$ the number of non-zero entries of a tensor.
The complexity of the first and the second term in the above formula are $O(\mathrm{nnz}(\tst{m}))$ and $O(C \cdot \ord{m} \cdot \mathrm{nnz}(\tst{m}) )$, respectively, thanks to the sparsity of $\tst{m}$.
With the complexity of the third term being $O(C^2 \cdot \sum_{i=1}^{\ord{m}} |\mc{V}_{\phi(m,i)}|)$, the overall complexity is reduced from $O(\prod_{i=1}^{\ord{m}} |\mc{V}_{\phi(m,i)}|)$ to $O(C \cdot \ord{m} \cdot \mathrm{nnz}(\tst{m}) + C^2 \cdot \sum_{i=1}^{\ord{m}} |\mc{V}_{\phi(m,i)}|)$.
That is, considering the previous example, the complexity of evaluating this Frobenius norm would decrease from a magnitude of $10^{20}$ to a magnitude of $10^{8}$. 

It is worth noting that the trick introduced in the last equivalency, Eq.~\eqref{eq::speed-up-obj}, has already been proposed in the study of Matricized Tensor Times Khatri-Rao Product (MTTKRP)~\cite{bader2007efficient, choi2014dfacto, smith2015splatt}.
MTTKRP and our model share a similarity in this trick because, unlike update rule Eq.~(5)\fakeref{} in the main file, evaluating the loss function Eq.~(3)\fakeref{} in the main file does not involve the non-negative constraints.

\section{Visualization of the Ablation Study}

\begin{figure}[ht]
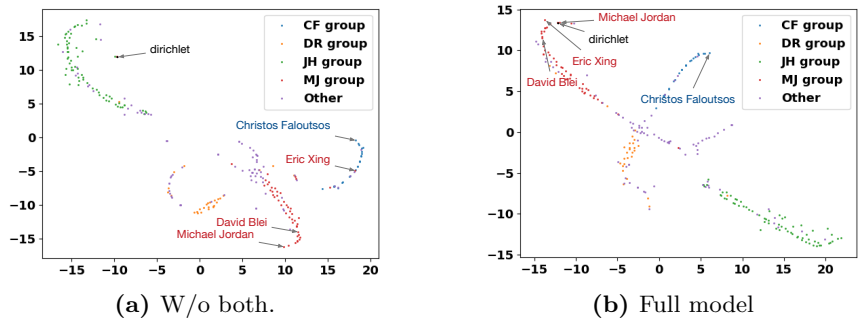

  \centering
  \begin{subfigure}[t]{.4\linewidth}
    \centering\includegraphics[width=\linewidth]{figures/eric_xing_vis_1}
    \caption{W/o both.}
  \end{subfigure}
  \qquad \qquad
  \begin{subfigure}[t]{.4\linewidth}
    \centering\includegraphics[width=\linewidth]{figures/eric_xing_vis_2}
    \caption{Full model}
  \end{subfigure}%\vspace{-6pt}
  \caption{Visualization of the ablation study where the author nodes are color-coded according to the truth label.}\label{fig::vis}
\end{figure}

To better understand the impact of candidate motif choice discussed in Section~7.3\fakeref of the main file, we further visualized the inferred membership of each node in Figure~\ref{fig::vis} by projecting its corresponding column in the consensus matrix $\matvcs{t}$ using t-Distributed Stochastic Neighbor Embedding (t-SNE).
As discussed in Section~4\fakeref, \textit{dirichlet} reflects a distinctive facet of the relationship between \textit{Xing} and \textit{Blei} pertaining to their graduating group.
The full model containing $AP4TPA$ inferred all of them to be close under the user guidance concerning research group.
In contrast, the partial model with only edge-level motifs not only mistakenly assigned \textit{Xing} to \textit{Faloutsos}'s group but also learned \textit{dirichlet} to be far away from either Xing or Blei.
This observation echos the intuition discussed in Section~4\fakeref that modeling higher-order interaction can introduce a richer pool of signals, and such modeling should be comprehensive and fine-grained in the task of user-guided clustering.

\begin{figure}[t]
  \centering
  \begin{subfigure}[t]{0.34\linewidth}
    \centering\includegraphics[width=\linewidth]{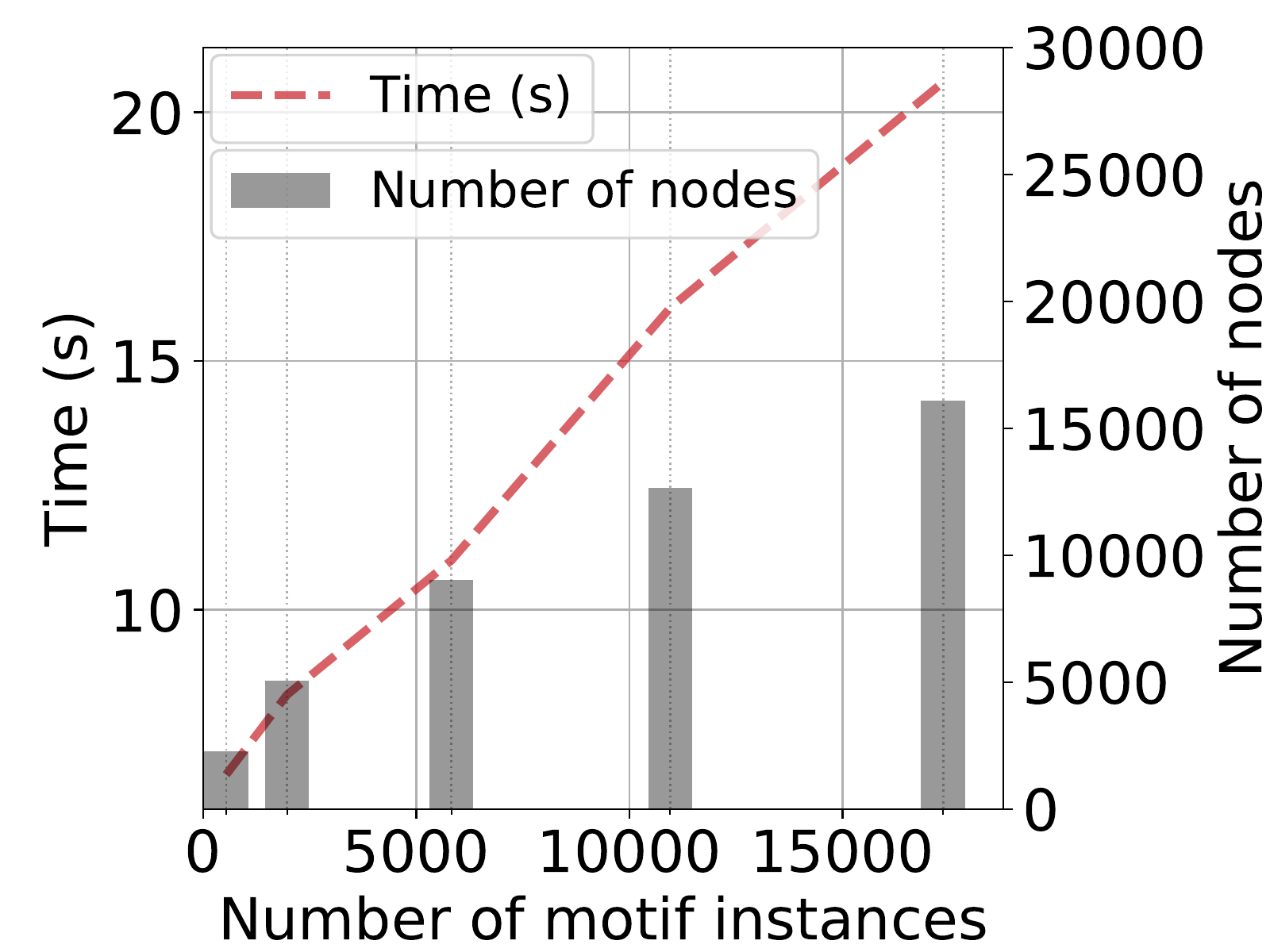}
    \caption{APPA in DBLP.}\label{fig::efficiency-appa}
  \end{subfigure}
  \begin{subfigure}[t]{0.34\linewidth}
    \centering\includegraphics[width=\linewidth]{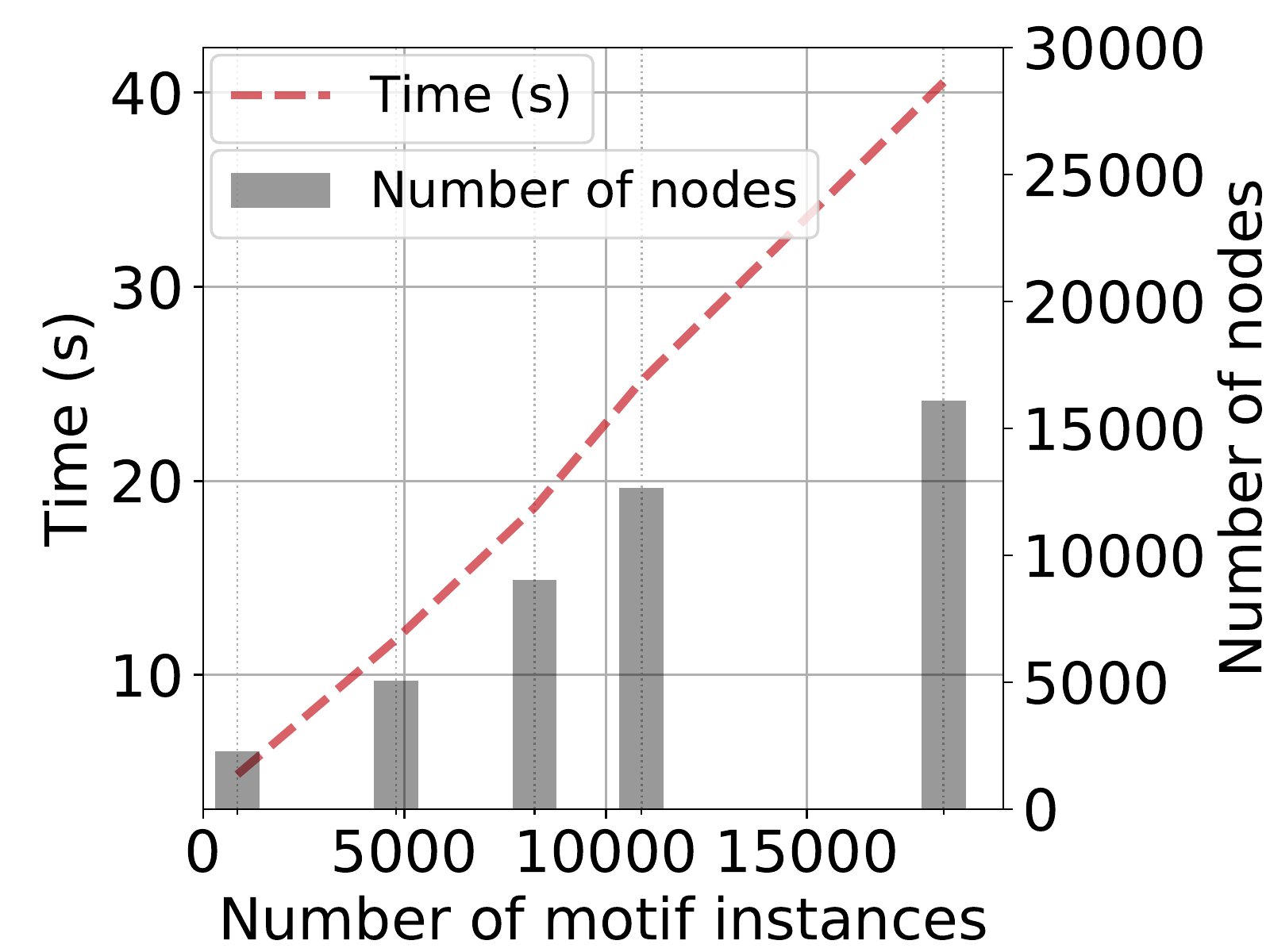}
    \caption{AP4TPA in DBLP.}\label{fig::efficiency-ap4tpa}
  \end{subfigure}
  \begin{subfigure}[t]{0.34\linewidth}
    \centering\includegraphics[width=\linewidth]{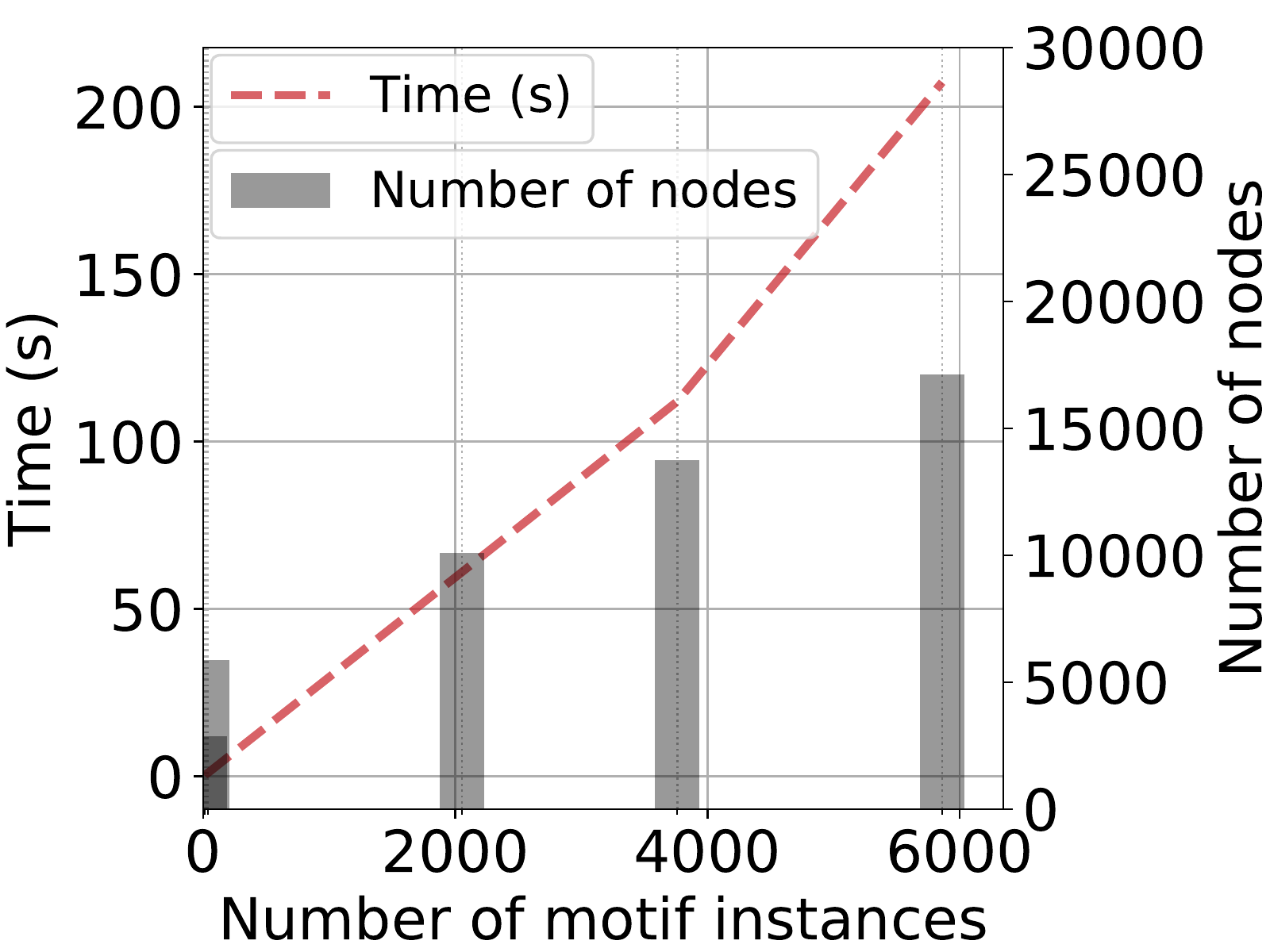}
    \caption{2P2W in YAGO.}\label{fig::efficiency-2p2w}
  \end{subfigure}
   \begin{subfigure}[t]{0.34\linewidth}
    \centering\includegraphics[width=\linewidth]{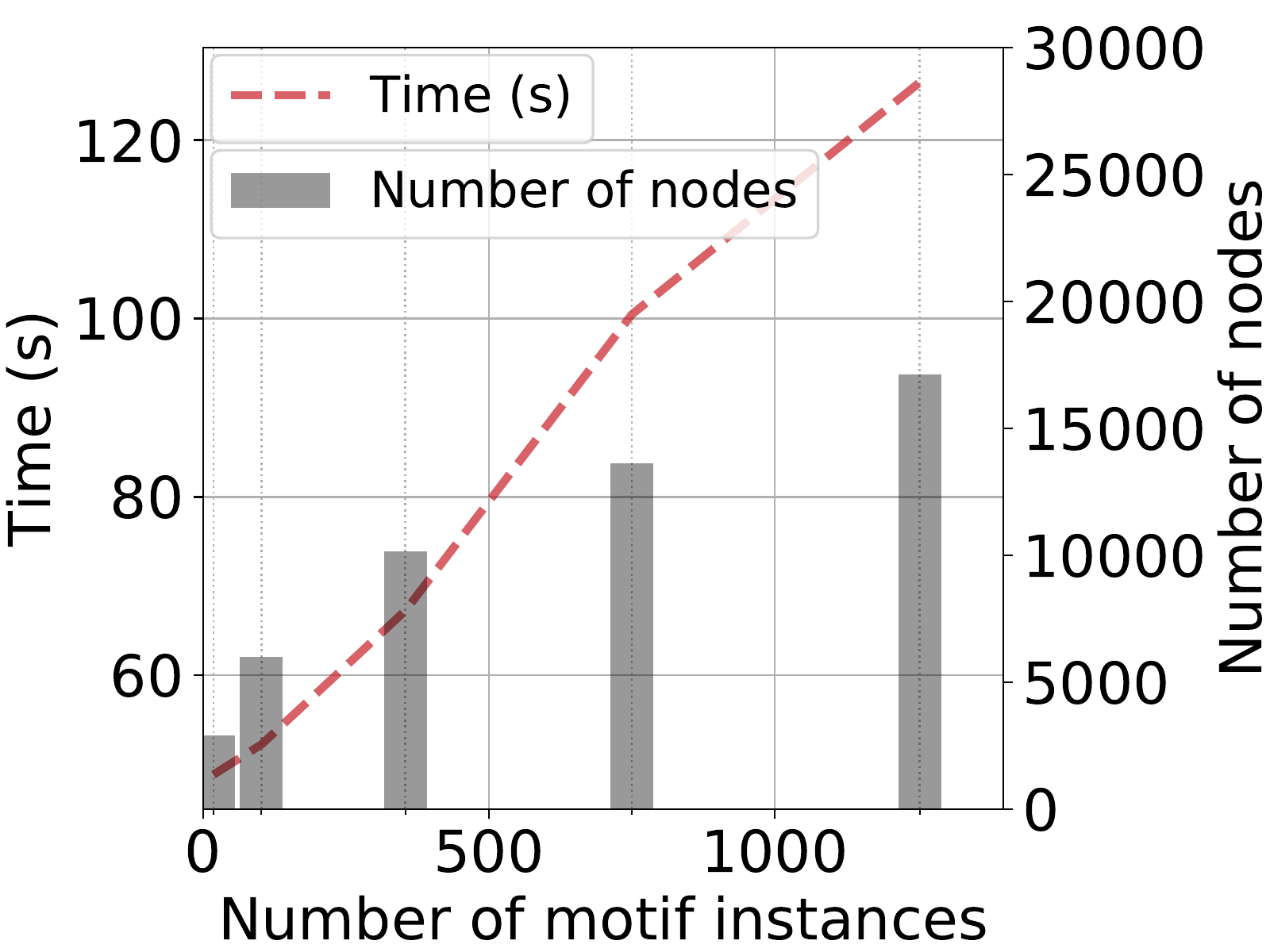}
    \caption{3PW in YAGO.}\label{fig::efficiency-3pw}
  \end{subfigure}
  \caption{Wall-clock runtime for inferring all parameters of one motif and the number of nodes against the number of motif instances in a series of downsampled HINs. The proposed algorithm empirically achieves near-linear efficiency, and motif instances are indeed sparse in HINs.}\label{fig::efficiency}
\end{figure}

\section{Efficiency Study}\label{sec::efficiency-study}

In this section, we empirically evaluate the efficiency of the proposed algorithm with a focus on the speed-up tricks described in Section~6\fakeref{} of the main file.
Specifically, we estimate the runtime for inferring all parameters involved in one motif while all other parameters are fixed, or equivalently, reaching convergence of the while-loop from line 4 to line 6 in Algorithm~1\fakeref{} in the main file.

This study was conducted on both the DBLP dataset and the YAGO dataset for each of their respective non--edge-level motifs: APPA and AP4TPA in DBLP; 2P2W and 3PW in YAGO.
The non--edge-level motifs are studied because (i) they are more complex in nature and (ii) the tensors induced by edge-level motifs are essentially matrices, the study of which degenerates to the well-studied case of non-negative matrix factorization.
To downsample the HINs, we randomly knock out a portion of papers in DBLP or persons in YAGO.
The involved edges and the nodes that become dangling after the knock-out are also removed from the network.
The reason node type paper and person are used is that they are associated with the most diverse edge types in DBLP and YAGO, respectively.
In the end, we obtain a series of HINs with $10\%$, $25\%$, $50\%$. $75\%$, $100\%$ of papers or persons left.

To more accurately evaluate the efficiency of the proposed algorithm in this study, we turn off the parallelization in our implementation and use only one thread.
We record the wall-clock runtime for inferring all parameters of each concerned motif, $\{\matv{l}{k}\}_{k = 1}^{\ord{l}}$, while fixing the motif weights $\vcmu$ and parameters of other motifs, $\{\matv{m}{i}\}_{m \neq l}$. 
The experiment is executed on a machine with Intel(R) Xeon(R) CPU E5-2680 v2 @ 2.80GHz.
The result is reported in Figure~\ref{fig::efficiency}.

\vpara{The proposed algorithm empirically achieves near-linear efficiency in inferring parameters of each given motif.}
As presented in Figure~\ref{fig::efficiency}, the runtime for all motifs on both datasets are approximately linear to the number of involved motif instances.
This result is in line with the analysis provided in Section~6\fakeref{} of the main file and justifies the effectiveness of the speed-up tricks.

Moreover, we also reported the number of motif instances against the number of nodes regardless of type in each downsampled network.
For all four studied motifs, we do observe motif instances are sparse and do not explode quickly as the size of the network increases.

\section{Detailed Description of Datasets, Evaluation Tasks, and Evaluation Metrics}
\begin{figure}[t]
  \centering
  \includegraphics[width=.8\linewidth]{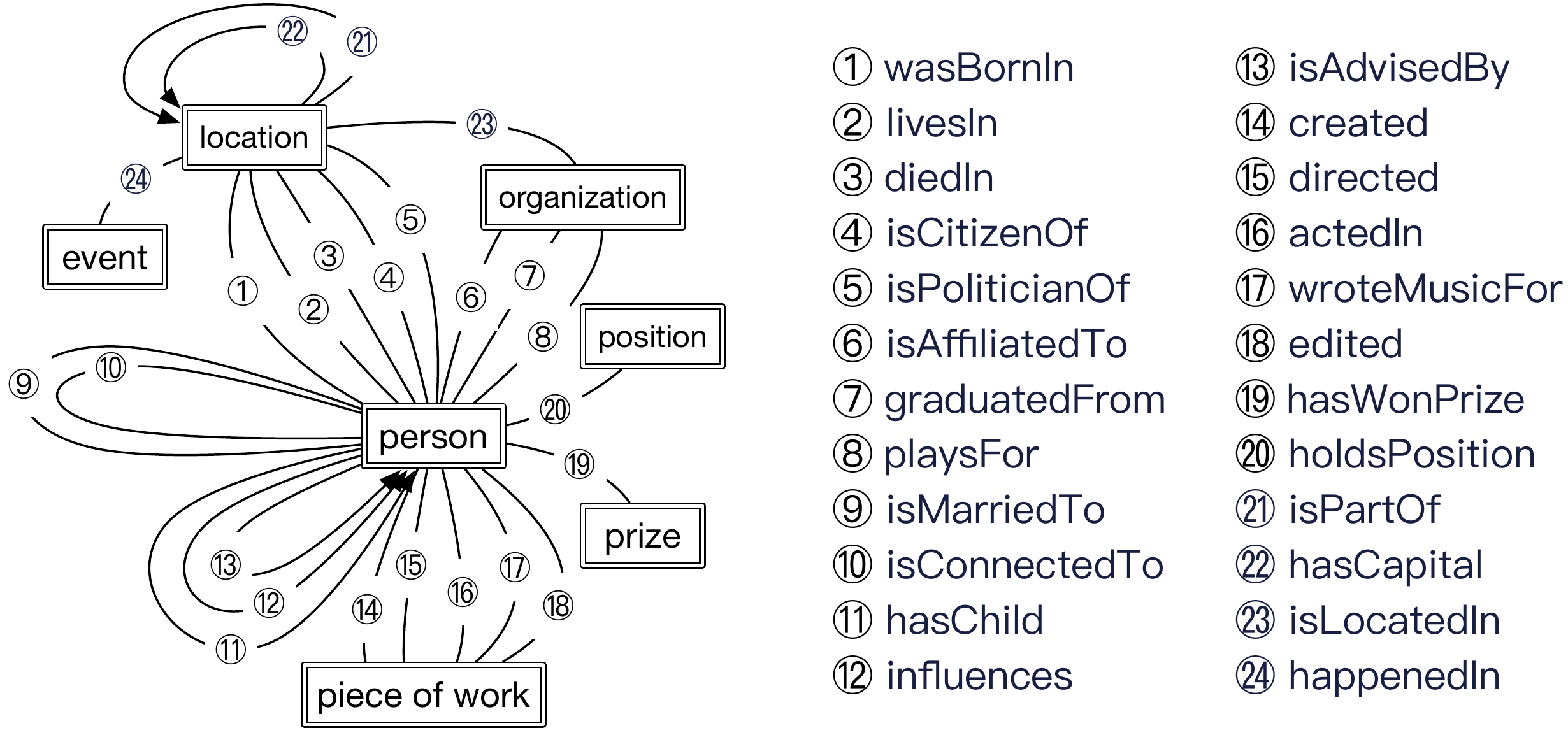}
    %\vspace{-9pt}
    \caption{The schema of YAGO~\cite{shi2018easing}.}\label{fig::yago-schema}
    %\vspace{-3pt}
\end{figure}

In this section, we provide the detailed description of the datasets, evaluation tasks, and metrics used in the experiments.
\vpara{Datasets.}
We use two real-world HINs for experiments.
\begin{itemize}
\item
\textbf{DBLP} is a heterogeneous information network that serves as a bibliography of research in computer science area~\cite{tang2008arnetminer}.
The network consists of 5 types of node: author ($A$), paper ($P$), key term ($T$), venue ($V$) and year ($Y$).
The key terms are extracted and released by Chen et al.~\cite{chen2017task}. 
The edge types include authorship, term usage, venue published, year published, and the reference relationship. 
The first four edge types are undirected, and the last one is directed. 
The schema of the DBLP network is shown in Figure~2a\fakeref{} in the main file.
In DBLP, we select two candidate motifs for all applicable methods, including $AP4TPA$ and $APPA$, where $APPA$ is also a meta-path representing author writes a paper that refers another paper written by another author and $AP4TPA$ was introduced in Section~3\fakeref{} of the main file.
\item
\textbf{YAGO} is a knowledge graph constructed by merging Wikipedia, GeoNames and WordNet. YAGO dataset consists of 7 types of nodes: person (\textit{P}), organization (\textit{O}), location (\textit{L}), prize (\textit{R}), work (\textit{W}), position (\textit{S}) and event (\textit{E}). 
There are 24 types of edges in the network, with 19 undirected edge types and 5 directed edge types as shown by the schema of the YAGO network in Figure~\ref{fig::yago-schema}.
In YAGO, the candidate motifs used by all compared methods include $P^{6}O^{23}L$, $P^{7}O^{23}L$, $P^{8}O^{23}L$, $2P2W$, $3PW$, where the first three are also meta-paths with the number in superscript being type of edge given in Figure~\ref{fig::yago-schema}.
$2P2W$ is the motif that 2 people simultaneously co-created (edge type $14$) two pieces of work, and $3PW$ is the motif that 3 people who created (edge type $14$), directed (edge type $15$) and acted (edge type $16$) in a piece of work, respectively.
\end{itemize}

\vpara{Evaluation tasks.}
In order to validate the proposed model's capability in reflecting different guidance given by different users, we use two sets of labels on authors to conduct two tasks in DBLP similar to previous study~\cite{sun2012integrating}.
Additionally, we design another task on YAGO with labels on persons.
We provide datasets and labels used in the experiment along with the submission.
%\begin{itemize}
%\item
\textbf{DBLP-group} -- 
Clustering authors to $5$ research groups where they graduated, which is an expanded label set from the ``four-group dataset''~\cite{sun2012integrating}. The ``four-group dataset'' includes researchers from four renowned research groups led by Christos Faloutsos, Michael I. Jordan, Jiawei Han, and Dan Roth. 
Additionally, we add another group of researchers, who have collaborated with at least one of the researchers in the ``four-group dataset'' and label them as the fifth group with the intention to involve more subtle semantics in the original HIN.
$5\%$ of the $250$ authors with labels are randomly selected as seeds from user guidance.
We did not use $1\%$ for seed ratio as in the following two tasks because the number of authors to be clustered in this task is small.
The resulted HIN processed as such consists of 19,500 nodes and 108,500 edges.
%\item
\textbf{DBLP-area} -- 
Clustering authors to $14$ research areas, which is expanded from the ``four-area dataset''~\cite{sun2012integrating}, where the definition of the 14 areas is derived from the Wikipedia page: List of computer science conferences\footnote{https://en.wikipedia.org/wiki/List   of   computer   science   conferences}. 
$1\%$ of the $7,165$ authors with labels are randomly selected as seeds from user guidance.
The HIN processed in this way has 16,100 nodes and 30,239 edges.
%\item
\textbf{YAGO} -- 
Clustering people to 10 popular countries in the YAGO dataset. 
We knock out all edges with edge type wasBornIn from the HIN. 
To generate ground truth labels, if a person is connected to one of the 10 countries through paths containing edges of type wasBornin, isPartOf, or has Capital, we assign this country to be the label of this person. 
Additionally, as happenedIn is not related to our task, it is removed as well.
$1\%$ of the $11,368$ people are randomly selected as seeds from user guidance.
There are 17,109 nodes and 70,251 edges in the processed HIN.
%\end{itemize}

\vpara{Evaluation metrics.}
We use three metrics to evaluate the quality of the clustering results generated by each model: Accuracy (Micro-F1), Macro-F1, and NMI.
\textbf{Accuracy} refers to a measure of statistical bias. More precisely it is defined by the division of the number of correctly labeled data by the total size of the dataset. 
Note that in multi-class classification tasks, accuracy is always identical to Micro-F1.
\textbf{Macro-F1} refers to the arithmetic mean of the F1 score across all different labels in the dataset, where the F1 score is the harmonic mean of precision and recall for a specific label. 
\textbf{NMI} is the abbreviation for normalized mutual information. Numerically, it is defined as the division of mutual information by the arithmetic mean of the entropy of each label in the data.
For all these metrics, higher values indicate better performance.

\section{Related Work on Matrix and Tensor Factorization for Clustering.}
By factorizing edges that represent pairwise interactions in a network, matrix factorization has been shown to be able to reveal the underlying composition of objects \cite{lee1999learning}.
In this direction, a large body of study has been carried out on clustering networks using non-negative matrix factorization (NMF)~\cite{liu2013multi, lee2001algorithms, ding2006orthogonal}.
As a natural extension beyond pairwise interaction, tensor has been used to model interaction among multiple objects for decades~\cite{tucker1966some, harshman1970foundations}.
A wide range of applications have also been discussed in the field of data mining and machine learning~\cite{papalexakis2017tensors, kolda2009tensor}.

For the study of clustering and related issues, many algorithms have been developed for homogeneous networks by factorizing a single tensor~\cite{shashua2005non, cao2016semi, sheikholeslami2016egonet, benson2015tensor, cao2015robust}. 
A line of work transforms a network to a $3$-rd order tensor via triangles, which is essentially one specific type of network motif~\cite{sheikholeslami2016egonet, benson2015tensor}.
Researchers have also explored weak supervision in guiding tensor factorization based analysis~\cite{cao2016semi}.
A large number of non-negative tensor factorization methods have been proposed for practical problems in computer vision~\cite{shashua2005non}.
Besides, tensor-based approximation algorithms for clustering also exist in the literature~\cite{sutskever2009modelling, cao2015robust}.  
One recent work on local network clustering considering higher-order conductance shares our intuition since it operates on tensor transcribed by a motif without decomposing into pairwise interactions~\cite{zhou2017local}.
This method is designed for the scenario where one motif is given.
Different from the approach proposed in our paper, all the above methods are not designed for heterogeneous information networks, where the use of multiple motifs is usually necessary to reflect the rich semantics in HINs.
Finally, we remark that to the best of our knowledge existing tensor-based clustering methods for HINs~\cite{gujral2018smacd, wu2017tensor} either do not jointly model multiple motifs or would essentially decompose the higher-order interactions into pairwise interactions.

\end{document}